\documentclass[manuscript,]{acmart}

\usepackage{amsmath, amssymb, amsfonts}
\usepackage{amsmath,amssymb,amsfonts}
\usepackage{algorithmic,algorithm}
\usepackage{graphicx,subcaption}
\usepackage{textcomp}

\AtBeginDocument{%
  }

\setcopyright{acmlicensed}
\copyrightyear{2026}
\acmYear{2026}
\acmDOI{XXXXXXX.XXXXXXX}

\acmJournal{JACM}
\acmVolume{37}
\acmNumber{4}
\acmArticle{111}
\acmMonth{8}

\begin{document}

\title{Trustworthy Second-hand Marketplace for Built Environment}

\author{Stanly Wilson}
\email{stanlywilson@stvincentngp.edu.in}
\orcid{0000-0002-1647-2251}
\affiliation{%
  \institution{Newcastle University}
  \city{Newcastle upon Tyne}
  \country{UK}
}

\affiliation{%
  \institution{St. Vincent Pallotti College of Engineering \& Technology}
  \city{Nagpur}
  \state{Maharashtra}
  \country{India}
}

\author{Kwabena Adu-Duodu}
\orcid{0000-0003-2418-9314}
\affiliation{%
  \institution{Newcastle University}
  \city{Newcastle upon Tyne}
  \country{UK}
}
\email{k.adu-duodu2@newcastle.ac.uk}

\author{Yinhao Li}
\orcid{0000-0001-6846-9161}
\affiliation{%
  \institution{Newcastle University}
  \city{Newcastle upon Tyne}
  \country{UK}
}
\email{Yinhao.Li@newcastle.ac.uk}

\author{Ringo Sham}
\orcid{0009-0006-5234-0615}
\affiliation{%
  \institution{Newcastle University}
  \city{Newcastle upon Tyne}
  \country{UK}
}
\email{Ringo.Sham@newcastle.ac.uk} 

\author{Yingli Wang }
\orcid{0000-0001-5630-9558}
\affiliation{%
  \institution{Cardiff University}
  \city{Cardiff}
  \country{UK}
}
\email{wangy14@cardiff.ac.uk}

\author{Ellis Solaiman}
\orcid{0000-0002-8346-7962}
\affiliation{%
  \institution{Newcastle University}
  \city{Newcastle upon Tyne}
  \country{UK}
}
\email{ellis.solaiman@newcastle.ac.uk}

\author{Charith Perera}
\orcid{0000-0002-0190-3346}
\affiliation{%
  \institution{Cardiff University}
  \city{Cardiff}
  \country{UK}
}
\email{pererac@cardiff.ac.uk}

\author{Rajiv Ranjan}
\orcid{0000-0002-6610-1328}
\affiliation{%
  \institution{Newcastle University}
  \city{Newcastle upon Tyne}
  \country{UK}
}
\email{Raj.Ranjan@newcastle.ac.uk}

\author{Omer Rana}
\orcid{0000-0003-3597-2646}
\affiliation{%
  \institution{Cardiff University}
  \city{Cardiff}
  \country{UK}
}
\email{RanaOF@cardiff.ac.uk}

\renewcommand{\shortauthors}{Wilson et al.}

\begin{abstract}
The construction industry faces significant challenges regarding material waste and sustainable practices, necessitating innovative solutions that integrate automation, traceability, and decentralised decision-making to enable efficient material reuse.
This paper presents a blockchain-enabled digital marketplace for sustainable construction material reuse, ensuring transparency and traceability using  InterPlanetary File System (IPFS). The proposed framework enhances trust and accountability in material exchange, addressing key challenges in industrial automation and circular supply chains. 
A framework has been developed to demonstrate the operational processes of the marketplace, illustrating its practical application and effectiveness. Our contributions show how the marketplace can facilitate the efficient and trustworthy exchange of reusable materials, representing a substantial step towards more sustainable construction practices. The evaluation is conducted on a prototype implementation under controlled conditions to analyse system behaviour and scalability trends, providing initial validation of the proposed approach.
\end{abstract}



\received{09 May 2025}
\received[revised]{22 April 2026}
\received[accepted]{5 July 2026}

\maketitle

\section{Introduction}

A trustworthy marketplace is a platform that ensures transparency, reliability, and fairness in every transaction, thereby building confidence among its participants. It provides clear, accessible information about products, prices, and transaction histories, minimising the risk of disputes and creating a uniform user experience \cite{IEEE_access_developing_trustworthy}. Trust in such marketplaces addresses common issues, such as fraud, misinformation, and inefficiencies, which often hinder stakeholder participation. By establishing mechanisms for transparency, accountability, security, and traceability, trustworthy marketplaces create an environment in which buyers feel confident in the authenticity and quality of products, and sellers are assured that their listings will reach genuine buyers without unwarranted conflicts. These factors collectively foster long-term engagement and a growing ecosystem of reliable trade \cite{reseller_market}.

Traditional marketplaces often rely on centralised authorities or intermediaries to mediate transactions, which introduces inefficiencies and potential bias \cite{Fame}. In contrast, a trustworthy marketplace incorporates mechanisms that ensure fairness and reduce dependency on intermediaries. Transparency is a cornerstone, allowing buyers and sellers to access detailed, verifiable product histories, which are often unavailable in traditional setups. Additionally, trustworthy marketplaces enforce accountability, enabling stakeholders to verify claims on products and services. These features make them more reliable and robust compared to traditional marketplaces, where trust often hinges on external oversight rather than intrinsic system design \cite{trust_e-marketplaces}.

The construction sector is the largest contributor to waste generation \cite{bbc_construction_waste}, and can benefit immensely from creating a marketplace that tracks the disposal of materials and facilitates their reuse. Many materials, such as steel, timber, and concrete, retain high value and can be reused, thereby reducing the demand for virgin resources and minimising environmental degradation. There are several barriers to reuse in the construction industry \cite{barriers_norway,barriers_enablers_sustainability}, including the lack of a marketplace, identification of reusable products, and trust related to product quality. 
Most existing systems do not adequately address the transition of materials from their first use to reuse in new construction projects. There is a clear need for a system that can track the quality, location, and availability of used materials, making them accessible for new projects and reducing reliance on new raw materials. This not only supports environmental sustainability but also enables economic savings in the construction industry.


A trustworthy marketplace can seamlessly work on blockchain technology, as blockchain provides a decentralised, transparent, and immutable system for recording transactions and verifying product authenticity. Blockchain addresses trust issues in marketplaces through three key mechanisms \cite{blockchain_exchange}: trust in exchange actors, actions, and assets. Traditional exchanges depend on human actors and centralised institutions for trust; however, blockchain replaces these with cryptography-driven protocols and eliminates reliance on intermediaries. Transparency and immutability ensure data integrity, while decentralised verification prevents tampering.  

Second-hand marketplaces using blockchains fundamentally differ from traditional marketplaces because of the inherent properties of blockchains, which offer several advantages. In conventional marketplaces, the administration of data and transactions is dependent on centralised systems. These systems are susceptible to fraud, lack transparency, and frequently involve intermediaries that raise costs and complicate dispute resolution. Second-hand marketplaces without using blockchains exist in areas such as batteries \cite{batteries_reuse}, the fashion industry \cite{cloth_reuse}, consumer electronics \cite{consumer_electronics_reuse}, and refurbished products \cite{refurbished}, which demonstrates the potential of second-hand marketplaces for the built environment. 
The proposed second-hand marketplace is not merely a commercial platform but supports sustainability initiatives, fostering the circular economy and supporting environmentally responsible building practices.

There are frameworks \cite{short_paper} that record the lifecycle of a product and its reuse; however, they have not clarified the criteria used to decide whether products are reusable or the details of how the product moves from the first user to the second user. The proposed trustworthy marketplace leverages the circular construction product ontology (CCPO) \cite{ccpo} to determine which products in the built environment are appropriate for reuse. The contributions of this study are as follows:
\begin{enumerate}
    \item We propose a second-hand marketplace for the built environment designed to enhance stakeholder trust and promote the sustainable reuse of construction materials.
    \item {We develop a domain-aware framework integrating digital product passport (DPP), blockchain, and IPFS to support verifiable product listing, reuse assessment, and secure transaction workflows.}
    \item {We present a prototype implementation and evaluation of the proposed framework, providing insights into system behaviour, scalability trends, and the effectiveness of application-level security mechanisms in a controlled environment.}
\end{enumerate}

The remainder of this paper is organised as follows. Section~\ref{sec: marketplace_literature} discusses relevant literature and related works. 
Section~\ref{sec:ch3_development} elaborately discusses the scenario under consideration, system development, and the underlying processes involved in the marketplaces. Section~\ref{sec: ch3_evaluation} evaluates the developed framework and its performance. Section~\ref{sec: ch3_conclusion} presents the conclusions and outlines future research directions.

\section{Literature and Related Works} \label{sec: marketplace_literature}
Various attempts have been made to use blockchains to enhance services associated with e-commerce. A blockchain-based framework \cite{acmdlt_fair_exchange}was developed to ensure fairness and transparency in digital content marketplaces, where facilitators mediate transactions between content owners and buyers. It addresses the challenge of passive content owners, who are typically excluded from the transaction process, by using smart contracts to ensure fair royalty distribution. Another framework \cite{ acmdlt_iot_reputaion} has been proposed to overcome the limitations of traditional centralised trust and reputation systems in IoT, addressing issues such as inefficiency, lack of scalability, and vulnerability to fraud. Li et al. \cite{TEM_luxuary} propose the use of blockchain to provide and verify the certificates of product quality to enhance trust in luxury e-commerce platforms. Their work explores the authentication process of certificates that verify products. This section explores the transformative potential of trustworthy marketplaces by examining the broader landscape of blockchain marketplaces, highlighting their ability to enhance transparency and efficiency. We then explore specific ways in which blockchains are integrated into the built environment. 

\subsection{Blockchain Marketplaces}
In recent years, various studies have used blockchains to bring trust and transparency to marketplaces. Liu et al. \cite{reseller_market} propose that integrating blockchain technology into supply chain management and e-commerce platforms can offer innovative insights for resolving traditional challenges within supply chain dynamics and recycling. The study underscores the critical role of blockchains in enhancing transparency and trust in recycling processes, increasing product recovery rates, and bolstering consumer confidence. Moreover, the study supports the use of blockchains for the reuse market, where increased trust and efficiency are essential for managing construction waste and facilitating material reuse.

Shi et al. \cite{awesome_conference} explored how blockchain-based decentralised marketplaces can overcome trust issues related to cloud service-level agreements (SLAs). They outlined the challenges associated with holding service providers accountable for the promised services. They propose the AWESOME framework implemented in Ethereum, which integrates smart contracts to automate and enforce service agreements, ensuring transparency and reliability. Ziyan et al. \cite{artchain} discuss the need for a trustworthy marketplace for the art industry, where fraud, lack of transparency, and provenance issues are rampant. They proposed the ArtChain framework, which is implemented on Ethereum and aims to leverage and provide traceable, tamper-proof ownership records and ensure secure transactions. 

Several data marketplaces have been proposed on the blockchain, such as quality data \cite{marketplaces_data_arxiv}, Internet of Things (IoT) data \cite{marketplaces_iotdata, iot_data_BIDM, fast_data}, and AI marketplaces \cite{AI_marketplace}. Michalopoulos et al. \cite{marketplaces_iotdata} presented a marketplace for IoT data by proposing a reputation and dispute resolution mechanism based on blockchain technology. The reputation system identifies and deals with tampering of reputation and ballot stuffing. However, it is unclear whether voters have access to data to identify the validity of a claim. The authors did not specify which blockchain platform the implementation is on but mentioned that Solidity is used to write smart contracts and there is no discussion on the implementation architecture or system design.

Banerjee et al. \cite{marketplaces_data_arxiv} highlight the increasing importance of data and point out the drawbacks of current data marketplaces. Their proposed design uses blockchain as a trusted third party, allowing for fair and transparent transactions and uses the exchange of medical data to illustrate the proposed marketplace design. However, it remains to be demonstrated whether it is practical to implement such a system at scale, and the paper does not provide any implementation details. 

Sarpatwar et al. \cite{AI_marketplace} discuss a blockchain implementation for an AI marketplace and propose a distributed protocol within a transfer learning setting where a consumer seeks to acquire a large dataset from various private data providers. However, the work showed performance degradation in its designs that provide higher trust guarantees and lacked details regarding the scalability of the solution. 

González et al. \cite{iot_data_BIDM} highlight the need for a trustworthy marketplace in the context of IoT data exchange, focusing on challenges like lack of transparency, centralization, and insufficient trust between stakeholders. They propose the Blockchain-based IoT Data Marketplace (BIDM), implemented on a permissioned blockchain, to enable secure, traceable, and transparent data transactions between producers and consumers.

Dixit et al. \cite{fast_data} highlight the importance of a trustworthy marketplace in ensuring secure and fair IoT data trading. Their proposed decentralised marketplace leverages blockchain technology, decentralised identity (DID), and peer-to-peer storage to enable transparent and tamper-proof transactions. The authors developed a proof-of-concept implementation using Hyperledger Fabric and VerneMQ.

 Unlike these data marketplaces, which trade intangible resources, the proposed marketplace facilitates the exchange of physical materials with a verifiable provenance.
 
\subsection{Integrating Blockchain for Built Environment}
Zhang et al. \cite{DPP_survey} provide a survey on the development of circular supply chain management and an elaborate list of projects/research related to the circular economy. They listed many projects that discussed circular supply chains, most of which identified the potential of blockchain technology; however, only a few dealt with real-time implementation and integration with blockchains. Yevu et al. \cite{digital_construction} study the importance of the digitalisation of the built environment and identify the potential of the blockchain to transform the construction supply chain.   

Treleaven et al. \cite{real_estate_marketplace} propose a blockchain-based real estate data marketplace to address challenges like data fragmentation, lack of transparency, and inefficiency in current systems. Their platform emphasises data tokenisation, validation, and accountability mechanisms to create a decentralised and trustworthy ecosystem. The authors recognise significant challenges in realising such a system, including handling unstructured data and diverse regulatory standards. However, the paper presents a conceptual framework and does not implement the proposed model.

Wilson et al.\cite{stanly_future_internet} have proposed a framework that monitors the lifecycle of a product in various stages of the supply chain and construction phases. The framework was built around a wood supply chain use case, and Polkadot and InterPlanetary File System (IPFS) were used to track and store various product information to build trust and transparency. Although this work provides significant insights, it fails to elaborate on the criteria for determining reuse and how products move from one building to another without any intermediary process.

Incorvaja et al. \cite{Omer_circular} discuss the circular supply chain for the built environment with a use case of LED light fittings. They implemented the framework in the Ethereum blockchain, which records information and ensures circularity. However, the scenario explains that the manufacturer took the products for inspection, repair, and reuse, which restricts the use of the product only with the manufacturer and not with the larger community. 

{Table~\ref{tab:comparison} presents a domain-wise comparison of existing blockchain-enabled marketplaces, traceability systems, and digital product passport implementations. Although previous works have explored trust and transparency across data marketplaces, e-commerce platforms, and circular product systems, most approaches either focus on intangible assets or lack mechanisms to support end-to-end reuse of physical materials. In the built environment, existing frameworks primarily address lifecycle tracking and provenance but do not integrate marketplace functionalities or structured reuse decision-making. Furthermore, none of the reviewed studies incorporate domain-specific ontologies to guide reuse eligibility. As shown in Table~\ref{tab:comparison}, the proposed framework uniquely combines blockchain-based traceability, DPP, marketplace-based exchange of physical materials, and ontology-driven reuse decision-making within a unified architecture.}
{Unlike existing studies that individually address DPP, provenance tracking, or blockchain-enabled marketplaces, the proposed framework integrates these capabilities within a single architecture specifically designed for construction material reuse. The incorporation of ontology-driven reuse assessment through CCPO further distinguishes the proposed approach by supporting systematic reuse decision-making prior to marketplace listing.}

\begin{table*}[htbp]
\centering
\caption{Comparison of Existing Approaches vs Proposed Framework}
\label{tab:comparison}

\resizebox{\textwidth}{!}{
\begin{tabular}{|l|l|c|c|c|c|c|c|}
\hline
\textbf{Work} & \textbf{Domain} & \textbf{Blockchain} & \textbf{DPP/ Traceability} & \textbf{Physical Asset Reuse} & \textbf{Marketplace} & \textbf{Ontology-based Decision} & \textbf{Implementation} \\ \hline

\cite{marketplaces_data_arxiv,fast_data,iot_data_BIDM,marketplaces_iotdata,AI_marketplace} 
                                                              & Data/IoT          & \checkmark & $\times$ & $\times$ & \checkmark & $\times$ & Prototype \\ \hline
\cite{acmdlt_fair_exchange,TEM_luxuary,awesome_conference,artchain} 
                                                              & Digital goods /  E-commerce    & \checkmark & $\times$ & $\times$ & \checkmark & $\times$ & Prototype \\ \hline

\cite{refurbished}                                            & Refurbished Products           & \checkmark & \checkmark  & \checkmark  & Limited & $\times$ & Conceptual \\ \hline
\cite{cloth_reuse,batteries_reuse,consumer_electronics_reuse}                                            
                                                              & Refurbished Products           & $\times$   & \checkmark  & \checkmark  & Limited & $\times$ & Conceptual \\ \hline

\cite{Omer_circular,stanly_future_internet}                   & Built Environment              & \checkmark & \checkmark  & \checkmark  &  $\times$  & $\times$ & Prototype \\ \hline

\cite{reseller_market,acmdlt_iot_reputaion}                   & Trust / Reputation Systems     & \checkmark & \checkmark  & Limited     &  $\times$  & $\times$ & Survey \\ \hline
\cite{digital_construction,DPP_survey}              & Built Environment              & $\times$   & \checkmark  & \checkmark  &  $\times$  & $\times$ & Survey \\ \hline

\cite{real_estate_marketplace}                                & Built Environment              & \checkmark & \checkmark  & \checkmark  & \checkmark & $\times$ & Conceptual \\ \hline
\textbf{This work}                                            & Built Environment              & \checkmark & \checkmark & \checkmark  & \checkmark   & \checkmark & Prototype \\ \hline

\end{tabular}
}
\vspace{2mm}
\footnotesize{\textit{Note:} “\checkmark” indicates full implementation, “$\times$ ” indicates no support, and “Limited” indicates component-level or partially integrated }

\end{table*}

\section{Motivation Scenario and Framework} \label{sec:ch3_development}
This section elaborates on the scenario and framework development with its various components and sequences of marketplace operations. To understand better, let us consider a use-case scenario from the construction sector. A building, several years post-construction, requires upgrades. Instead of sending directly to recycling or to a landfill, building owners prefer to reuse/repurpose the building’s high-quality components. However, there are a few challenges to be addressed. How can they ensure the reuse of their products in some other buildings/projects without advertising the availability of such materials? How can buyers be sure about the quality of the products if they buy them from their current owners? How can sellers be responsible for the quality of their products? The proposed framework addresses these challenges by allowing building managers to identify what products can be reused and list reusable components on a dedicated, blockchain-enabled marketplace. The framework generates a digital product passport (DPP) \cite{DPP_survey} for each product with its specifications, stores them on the IPFS, and logs this information securely on the blockchain, ensuring that all items listed are verifiable and traceable. This guarantees the materials' origin and provides a transparent transaction history, thereby facilitating auditability and enhancing trust. 
\subsection{System Framework}
\begin{figure}[!ht]
\centering
\centerline{\includegraphics[scale=0.67]{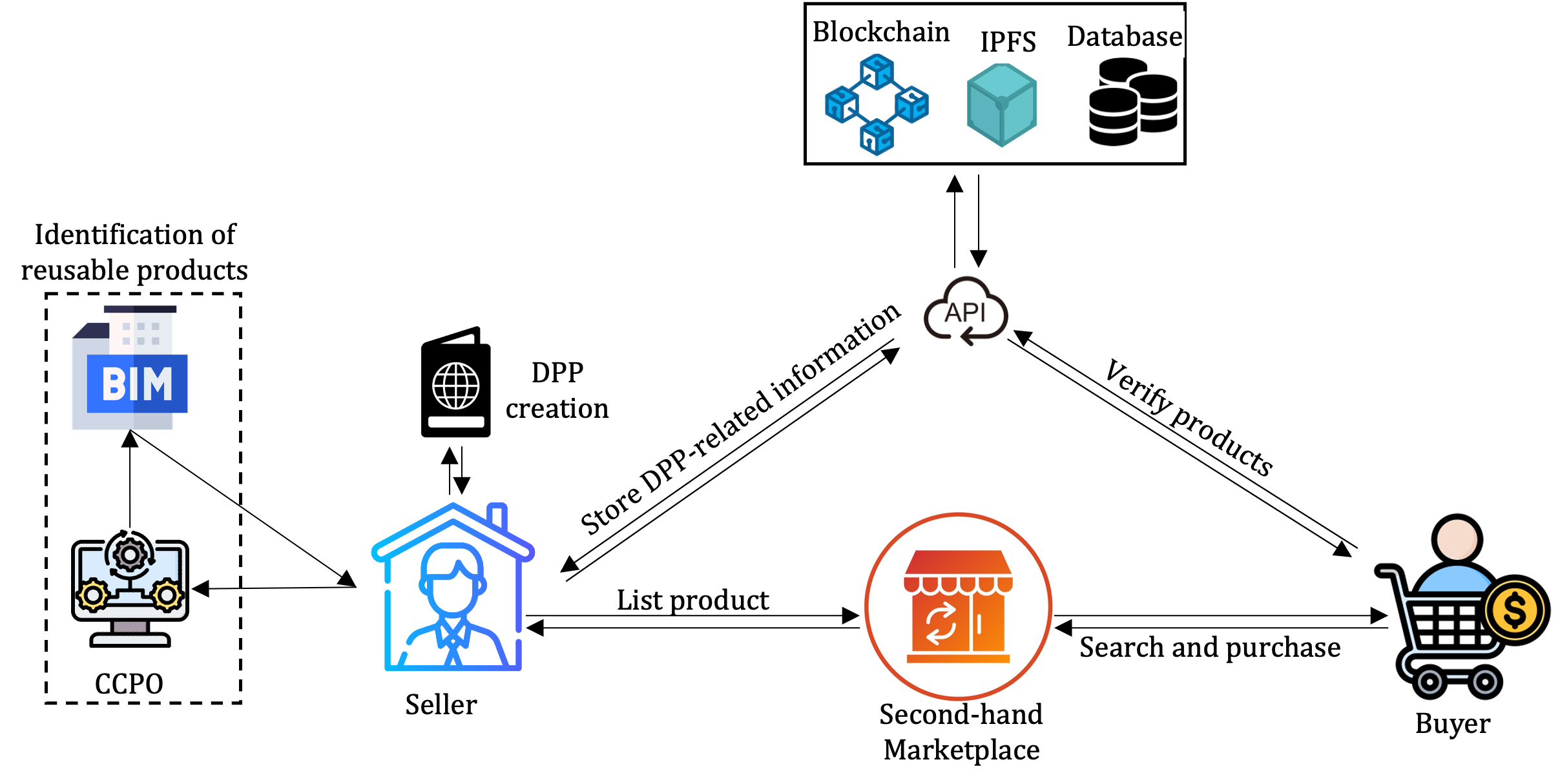}}
\caption{Trustworthy Marketplace Framework}
\Description{Trustworthy Marketplace Framework}
\label{architecture}
\end{figure}

The system is designed to facilitate a seamless and secure marketplace environment. Figure~\ref{architecture} shows the framework of the proposed trustworthy marketplace with various entities, such as sellers, buyers, storage, building information modelling (BIM), CCPO, and DPP modules. The sellers (those who are permitted to add products to the marketplace) using the CCPO framework access the BIM to identify reusable items. After identifying the reusable items, a DPP for the same is created, which is then pushed to the IPFS and the blockchain and stores other less critical information on the off-chain databases for accessing the marketplace.{The DPP encapsulates key attributes of the product such as product identity, material composition, condition, lifecycle history, and ownership records, enabling comprehensive traceability and verification throughout the reuse process.} Interested buyers can find these products on the marketplace and verify their authenticity and provenance on blockchain and IPFS. The detailed flow of the process and framework implementation are discussed in the following sections. 

\begin{figure}[!ht]
\centering
\includegraphics[scale=0.7, trim=0.1cm 0.05cm 0.05cm 0.05cm, clip]{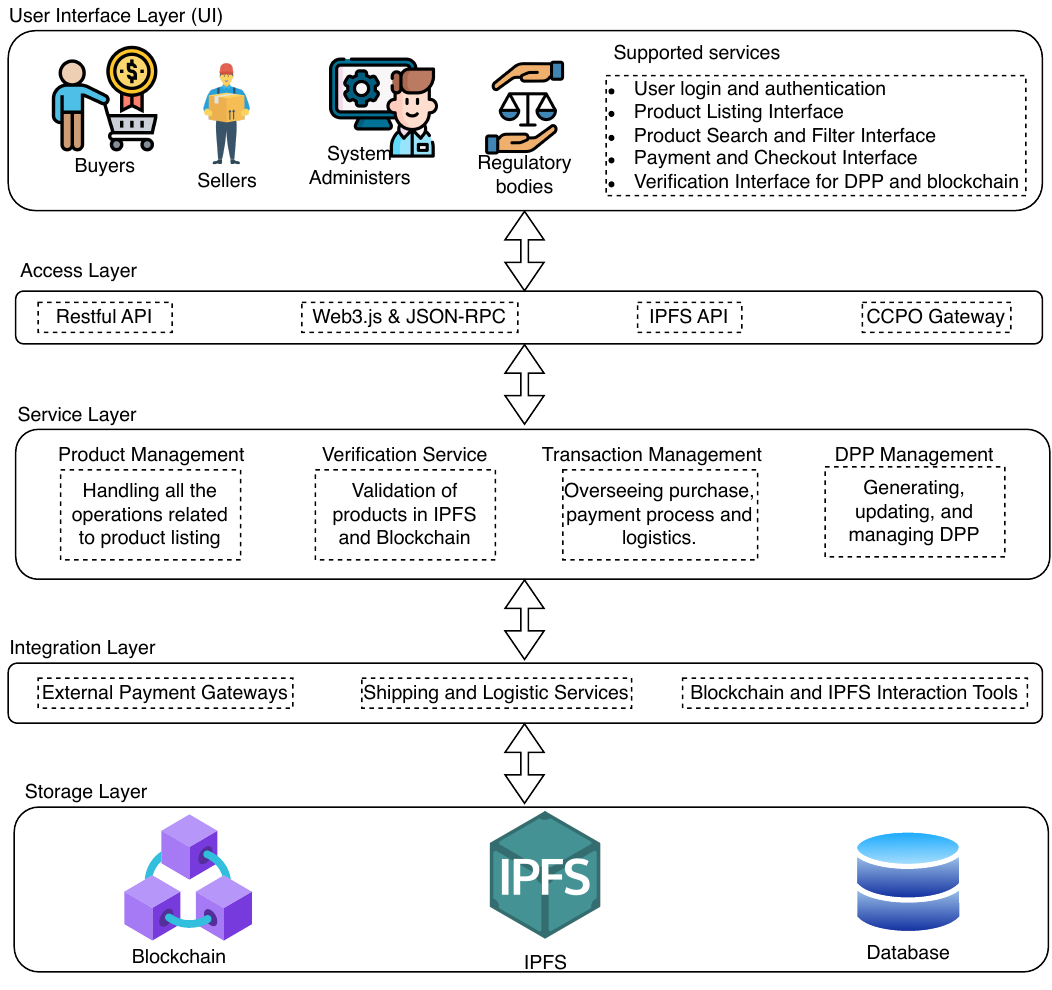}
\caption{System Framework}
\Description{System framework of the proposed trustworthy marketplace}
\label{layer_architecture}
\end{figure}

The system framework is meticulously designed to facilitate a seamless and secure marketplace environment using a multi-layered architecture. Each layer plays a crucial role in ensuring that the entire ecosystem functions efficiently and securely, from the user interface to the storage systems. Figure~\ref{layer_architecture} shows the system framework of the proposed marketplace.

The user interface (UI) layer serves as the front-end where interactions between the marketplace and its users take place. It is equipped with comprehensive services that enhance user experience and operational efficiency, including user login and authentication, ensuring secure access to the marketplace. Additionally, the product listing service allows sellers to showcase their items effectively, while the product search and filter interface provides buyers with a powerful tool to locate and select products based on specific criteria. The payment and checkout interface facilitates smooth transaction processes, and a verification interface for DPP and blockchain ensures that all product information is verified and trustworthy, maintaining the integrity of transactions.

The access layer is a critical gateway for all data requests entering and leaving the system. The RESTful API provides a standard method for external systems to communicate with the marketplace. Web3.js and JSON-RPC allow the marketplace to interact directly with the blockchain, enabling smart contract integrations and direct blockchain queries. The IPFS API is crucial for handling decentralised file storage, ensuring that DPPs are securely managed and accessible. The CCPO gateway manages the interaction between the marketplace and the CCPO interface. A basic explanation of how the CCPO works is provided in Section ~\ref{sec: market_ccpo}. 

The service layer is the operational backbone of the marketplace, which includes product management that oversees all aspects of product listings, from creation to updates. The validation service confirms the authenticity and integrity of products listed on the marketplace by verifying them against IPFS and blockchain records. Transaction management orchestrates the purchase process, from payment handling to overseeing logistics, ensuring that transactions are executed smoothly and efficiently. DPP management generates, updates, and maintains DPPs, ensuring that they are accurate and reflect current product ownership and details.

The integration layer handles all interactions with external systems, such as integrating external payment gateways that facilitate financial transactions, shipping and logistics services that manage the delivery of physical goods, and tools for the seamless integration of blockchain and IPFS. 

The storage layer houses data storage components, such as blockchains, IPFS, and traditional databases. The blockchain component records all transactional data, ensuring that it is immutable and transparent. {On the blockchain, the smart contracts implement core functions such as role authentication, product registration, ownership transfer, and verification of transactions, ensuring that all operations are executed according to predefined rules and recorded immutably.} The framework uses Polkadot, a heterogeneous multichain platform built on Substrate. IPFS stores immutable product details, enhances security, and provides a reliable means of accessing product information. Databases store user information, transaction records, and other operational data necessary for smooth functioning of the marketplace, ensuring that the data are readily available and securely managed.

There are various interactions take place in the system and to better understand user engagement, we present four critical interactions: CCPO process, product listing, verification, and buyer-seller interaction and payment.
\subsubsection{CCPO Process} \label{sec: market_ccpo}
CCPO helps the building managers/sellers to identify which products can be considered for reuse. The detailed version can be found in \cite{ccpo}. Various elements and their relationships are detailed below:
\begin{itemize}
    \item \( P \): Set of all products available in the building.
    \item \( V \): Set of value fields for each product, providing specific details such as material, condition, and usage history.
    \item \( R \): Reuse category for products:
    \begin{itemize}
        \item \( R_{strong} \): Strong reuse category where the product can be directly listed.
        \item \( R_{weak} \): Weak reuse category where repairs are needed.
        \item \( R_{none} \): No reuse potential; product should be recycled or moved to landfill.
    \end{itemize}
    \item \( P_{j} \): Product \( j \), where \( j \in (1, M) \), and \( M \) is the total number of products in the building.
    \item \( V_{j} \): Set of values associated with product \( P_{j} \), which includes attributes like material, age, condition, and estimated lifecycle.
    \item \( R_{j} \): Reuse category for product \( P_{j} \).
\begin{equation}
    R_{j} = 
    \begin{cases}
    R_{strong} & \text{if } f(V_{j}) \geq \theta_{strong}, \\
    R_{weak} & \text{if } \theta_{weak} \leq f(V_{j}) < \theta_{strong}, \\
    R_{none} & \text{if } f(V_{j}) < \theta_{weak.}
    \end{cases}
\end{equation}
    \item \( f(V_{j}) \): Evaluation function that uses the value fields to determine the product's potential for reuse.
    \item \( \theta_{strong} \) and \( \theta_{weak} \): Thresholds to determine strong reuse, weak reuse, or no reuse potential.
\begin{equation}
    L_{j} = 
    \begin{cases}
    1 & \text{if } R_{j} = R_{strong}, \\
    0 & \text{otherwise.}
    \end{cases}
\end{equation}
\item \( L_{j} \) represents the listing status of the product \( P_{j} \) and this function decides whether the product can be listed in the marketplace. If \( R_{j} \) is categorised as strong reuse, the product is approved for listing (\( L_{j} = 1 \)).
\item \( A_{j} \) represents the action taken for product \( P_{j} \). If a product falls under \( R_{weak} \) or \( R_{none} \), further action is taken based on the reuse assessment. Weak reuse products are repaired to improve their reusability, whereas non-reusable products are either recycled or moved to a landfill.
\end{itemize}
\begin{equation}
    A_{j} = 
    \begin{cases}
    \text{Repair} & \text{if } R_{j} = R_{weak}, \\
    \text{Recycle or Landfill} & \text{if } R_{j} = R_{none.}
    \end{cases}
\end{equation}
\subsubsection{Product Listing}
It focuses on the interaction between building managers (acting as sellers), buyers, and the marketplace system when listing construction materials. Various elements and their relationships are detailed below:
%
%
    \begin{itemize}
        \item $P$ = set of all products available for listing.
        \item $M$ = set of building managers overseeing materials.
        \item $L$ = set of all listings made by building managers.
        \item $V$ = verification process for materials listed on the marketplace.
        \item $PID_i$ = product ID for product $i$, where $i \in (1, P)$.
        \item $MID_j$ = building manager ID (seller) for manager $j$, where $j \in (1, M)$.
        \item $LID_{ij}$ = listing ID of product $i$ listed by manager $j$.
    \end{itemize}
The following equations represent whether a product listing is valid and verified before being made available to buyers:
    \begin{equation}
    X = \begin{cases}
    1,  \text{if } PID_i \in P \text{ and } MID_j \in M, \\
    0, \text{otherwise.}
    \end{cases}
    \end{equation}
    \begin{equation}
    Y = \begin{cases}
    1,  \text{if } LID_{ij} \in L \text{ and verified,} \\
    0, \text{otherwise.}
    \end{cases}
    \end{equation}
The verification value $V$ depends on $X$ and $Y$. $V = X \times Y$ must be equal to 1 to confirm that a product is credibly listed and verified. The operations of adding the product to the marketplace can be found in section~\ref{sec: market_add_prodcut}. The verification referred to in this section does not involve IPFS and blockchain.  
\subsubsection{Verification on IPFS and Blockchain}
It focuses on the verification of DPP on IPFS and blockchain. It evaluates the consistency and validity of DPP for the listed products. The elements, relationships, and respective objective functions are as follows:
\begin{itemize}
    \item \(DPP\): Digital Product Passport containing all product details stored in IPFS.
    \item \(IPFS_{ref}\): IPFS reference (hash) of the DPP stored in IPFS.
    \item \(BC_{ref}\): Blockchain reference that stores the \(IPFS_{ref}\), ensuring traceability.
    \item \(V\): Verification status to determine if the product information is valid.
    \item \(P_i\): Product \(i\), where \(i \in (1, N)\) and \(N\) is the total number of products listed.
    \item \(IPFS_{ref_i}\): IPFS reference for product \(i\)'s DPP.
    \item \(BC_{ref_i}\): Blockchain reference pointing to the \(IPFS_{ref_i}\).
\end{itemize}
\begin{equation} \label{equ:market_valid}
    X_i = 
    \begin{cases} 
    1 & \text{if } P_i \in \text{Marketplace and } Y_i = 1, \\
    0 & \text{otherwise.}
    \end{cases}
\end{equation}
The equation~\ref{equ:market_valid} checks if the product \(P_i\) exists in the marketplace and that the DPP reference is valid.
\begin{equation}\label{equ:market_block_valid}
    Y_i = 
    \begin{cases} 
    1 & \text{if } IPFS_{ref_i} = BC_{ref_i}, \\
    0 & \text{otherwise.}
    \end{cases}
\end{equation}
If \(Y_i = 1\), it implies that the DPP stored on IPFS is the original version referenced by the blockchain, maintaining immutability.
\begin{equation}
    V_i = X_i \times Y_i.
\end{equation}
If \(V_i = 1\), it indicates that the product’s DPP is valid, the IPFS reference matches the blockchain, and the product is properly listed in the marketplace. The detailed verification process can be found in section~\ref{sec: market_veify_product}
\subsubsection{Buyer-Seller Interaction and Payment}
It involves buyers browsing available materials, purchasing decisions, and the subsequent payment process. Various elements and their relationships are detailed below:
%
%
\begin{itemize}
    \item $B$ = set of buyers interacting with the marketplace, 
    \item $O$ = set of orders placed, 
    \item $T$ = transaction status, detailing whether a payment has been successfully completed, 
    \item $L$ = set of all product listings made by sellers, 
    \item $BID_m$ = buyer ID of buyer $m$, where $m \in (1, B)$,  
    \item $LID_n$ =  listing ID for the product $n$, where $n \in (1, L)$, 
    \item $OID_{mn}$ = order ID for the product placed by buyer $m$ for listing $n$, 
    \item $TID_p$ = transaction ID for order $OID_{mn}$ confirming payment, where $p \in (1, T)$.
\end{itemize}
To ensure the integrity of transactions between buyers and sellers, the following objective functions are defined:
 \begin{equation}
    X = \begin{cases}
    1,  \text{if } BID_{m}\in B, \\
    0,  \text{otherwise.}
    \end{cases}
    \end{equation}
   \begin{equation}
    Y = \begin{cases}
    1,  \text{if } OID_{mn} \in O  \text{ and relates to $LID_n$}, \\
    0,  \text{otherwise.}
    \end{cases}
    \end{equation}
    \begin{equation}
    Z = \begin{cases}
    1,  \text{if } TID_{p} \in T  \text{ and payment is successful}, \\
    0,  \text{otherwise.}
    \end{cases}
    \end{equation}
The overall transaction status ($T_{status}$) depends on X, Y and Z.
\begin{equation}
    T_{status} = X \times Y \times Z.
\end{equation}
The transaction is considered successful if $T_{status}$ = 1, ensuring that the buyer is valid, the order is appropriately placed, and payment has been processed successfully. Further elaboration of the payment process will be discussed in section~\ref{sec: market_payment}.

\subsection{System Operations} \label{sec: market_operations}
\begin{figure}
\centering
  \includegraphics[scale=0.70, trim=0.3cm 9.05cm 0.3cm 8.75cm, clip]{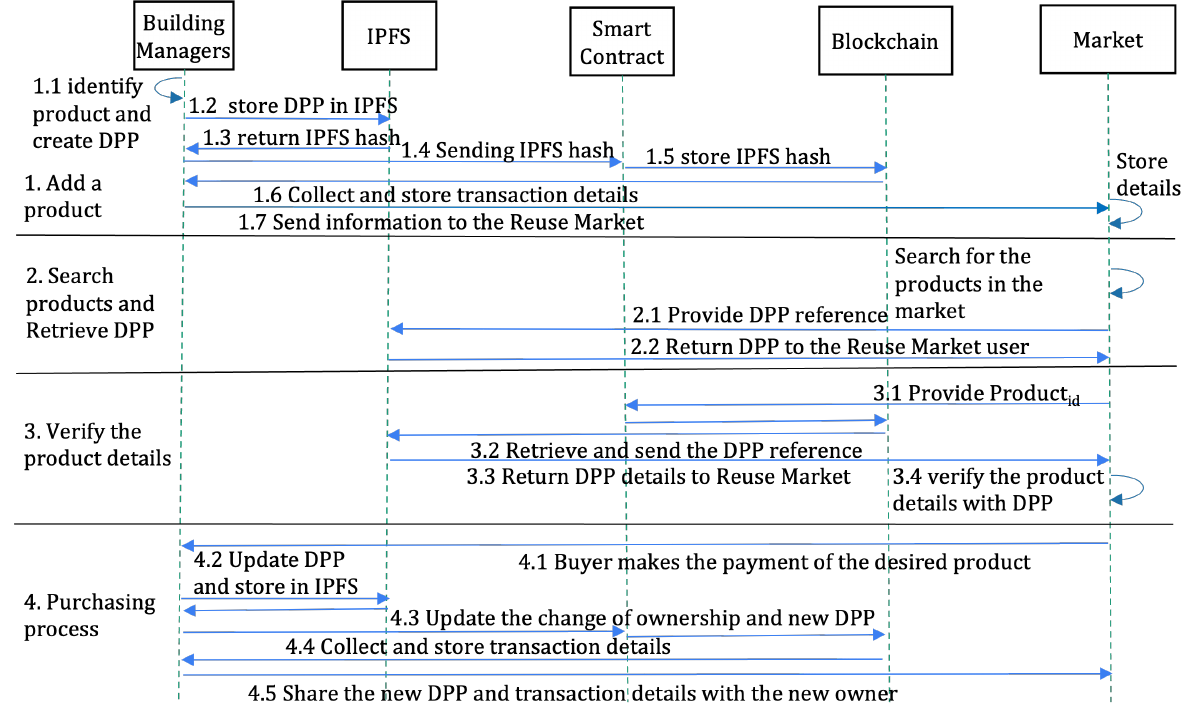}
  \caption{Sequence Diagram for Marketplace Operations}
  \label{sequence}
  \Description{Sequence diagram for the proposed marketplace operations}
\end{figure}

Having discussed the system framework, it is appropriate to elaborate on the four underlying operations of the marketplace, and Figure \ref{sequence} shows a sequence diagram of the same. For better clarity, the individuals or entities who make decisions regarding product reusability and bring the product to the marketplace are referred to as managers or sellers. 
{The algorithms presented in this section are designed as domain-aware workflows tailored to the construction reuse context. Unlike generic marketplace operations, these workflows integrate reuse eligibility assessment (via CCPO), lifecycle-aware data representation (via DPP), and verifiable provenance tracking (via blockchain and IPFS). The design reflects the unique challenges of handling heterogeneous physical assets, ensuring data integrity across distributed systems, and maintaining trust during ownership transitions.}

\subsubsection{Add Product} \label{sec: market_add_prodcut}
Based on the CCPO framework, building managers can determine which product can go to the reusable market based on various metrics. During renovation or repurposing, they can push the desired products that meet the reusable criteria to the marketplace. This process of moving products to the market can be manual or automatic based on the conditions set. When a product is identified as reusable, a DPP must be created using all available information regarding the product. The details in the DPP may vary from product to product. As no standard specifies the information that must be contained in a DPP, we aim to keep general information such as product identification, material composition, design and manufacturing, usage information, maintenance and repair, regulatory compliance, and end-of-life details. As the DPP evolves, more information may need to be added to this list. DPPs may also contain text, diagrams, and images.

\begin{algorithm}[H]
\caption{Adding a product to marketplace}
\label{algo: add_product}
\begin{algorithmic}[1]
\STATE $DPP \leftarrow \text{CreateDPP}(p)$
\STATE $A_{IPFS} \leftarrow \text{StoreInIPFS}(DPP)$
\STATE $Info \leftarrow {(A_{IPFS}, CPD)}$
\STATE $T_{id} \leftarrow \text{RecordOnBlockchain}(Info)$
\STATE $\text{UpdateMarketplace}(A_{IPFS}, T_{id},CPD)$
\end{algorithmic}
\end{algorithm}

Algorithm~\ref{algo: add_product} details the process of adding a product to the marketplace. It begins by generating a DPP for a given product \textit{p\textsubscript{id}} with various available details. The DPP is then stored in IPFS, the IPFS address (\textit{A\textsubscript{IPFS}}) and critical product details (\textit{CPD}) are packaged into an information set (\textit{Info}), which is then recorded on the blockchain. If the blockchain recording transaction fails (that is, \textit{T\textsubscript{id}} is null), the recording step is repeated. A basic assumption is that every product has a unique identifier on the blockchain, and the system throws an error if anyone attempts to add the same product again; however, updates are permitted to the added product. A successful execution of these steps returns \textit{T\textsubscript{id}} and effectively adds the product to the marketplace with verifiable and immutable records such as \textit{A\textsubscript{IPFS}}, \textit{T\textsubscript{id}}, and \textit{CPD}. {Unlike conventional listing mechanisms, this process is preceded by reuse eligibility assessment using CCPO, ensuring that only materials classified under strong reuse are listed. Furthermore, the DPP is generated from heterogeneous construction data sources, which introduces domain-specific complexity that is not present in generic marketplaces. The coupling of IPFS storage with blockchain references ensures immutable provenance, which is critical for verifying the history of physical construction materials.}

\subsubsection{Search Product and Retrieve DPP} \label{sec: market_retrieve_prodcut}
The marketplace interface is designed to enhance user engagement through advanced search capabilities, allowing precise filtering based on specific attributes and user preferences. This functionality allows users to effectively narrow down their choices based on particular preferences, displaying key product attributes and essential details directly within marketplace listings. Upon identifying a product of interest, users can access the DPP from the IPFS to learn a comprehensive product profile detailing its manufacturing history, material composition, usage guidelines, and compliance with regulatory standards. In this way, the DPP serves not only as a testament to the product's quality and compliance but also enhances consumer trust by ensuring transparency and accessibility of vital product information.

\subsubsection{Verification Process} \label{sec: market_veify_product}
The framework employs a robust combination of IPFS and blockchain to enhance verification and ensure provenance integrity. IPFS offers the versatility of storing diverse file types; however, managers can update and revise the DPP and offer a different DPP to the buyer other than what was listed in the marketplace with malicious intentions. The blockchain component plays a crucial role in safeguarding against potential discrepancies. It verifies the authenticity of the DPPs by maintaining a reference on the blockchain, ensuring that the document presented by the seller matches the one recorded in the marketplace. This mechanism maintains transparency and trust between sellers and buyers. Additionally, this verification process allows buyers to compare the pricing details stored on the blockchain with the marketplace listing, ensuring no discrepancies in the charges for the product and fostering a fair-trade environment.

\begin{algorithm}[H]
\caption{Verification Process}
\label{algo: verification_process}
\begin{algorithmic}[1]
\STATE $DPP_{marketplace} \leftarrow \text{GetMarketPlaceDPP}(p_{id})$
\STATE $DPP_{blockchain} \leftarrow \text{GetDPP}(T_{id},p_{id})$
\IF {$DPP_{blockchain} \neq DPP_{marketplace}$}
    \RETURN "Discrepancy found: DPP does not match."
\ELSE
    \STATE $Price_{blockchain} \leftarrow \text{GetBlockchainPrice}(T_{id}, p_{id})$
    \STATE $Price_{marketplace} \leftarrow \text{GetMarketplacePrice}(p_{id})$
    \IF {$Price_{blockchain} \neq Price_{marketplace}$}
        \RETURN "Verification failed: Pricing does not match."
    \ELSE
         \RETURN "Verification successful."
    \ENDIF
\ENDIF
\end{algorithmic}
\end{algorithm}

Algorithm~\ref{algo: verification_process} describes the underlying verification process. Let \textit{p\textsubscript{id}} be a product with a blockchain transaction details \textit{T\textsubscript{id}}. Using \textit{p\textsubscript{id}} and \textit{T\textsubscript{id}}, interested parties can verify the product by comparing the information available in the marketplace and the information stored on the blockchain. The verification process first compares the DPPs stored in the marketplace and on the blockchain. If they are the same, the comparison is made of the product price listed on the marketplace and the blockchain. If the comparison is true, the verification is successful; otherwise, it is unsuccessful. {This verification mechanism is particularly important in the context of construction material reuse, where product information may evolve over time and may be intentionally or unintentionally altered after listing. Unlike digital goods, physical assets are subject to condition changes, incomplete documentation, and asymmetric information between buyers and sellers. By enforcing consistency between marketplace data and blockchain-referenced IPFS records, the framework ensures that the product information presented to buyers is authentic and has not been tampered with. Additionally, the inclusion of price verification helps prevent marketplace-level manipulation, thereby strengthening trust across all stakeholders.}

\subsubsection{Payment Process} \label{sec: market_payment}
In addition to listing, searching, and verifying products, our marketplace framework facilitates a streamlined purchasing process. Due to the limited mainstream acceptance and various regulatory environments surrounding cryptocurrency, our platform has chosen to integrate with established payment methods rather than adopting crypto-based payments. Nonetheless, our framework is compatible with integrating blockchain-based payment solutions, as discussed in \cite{marketplaces_data_arxiv, payment_polkadot}. 
This approach ensures a familiar and secure payment experience for users. Buyers can proceed to the purchasing phase once a product has been selected and its authenticity verified through the DPP and IPFS. During this stage, financial transactions are conducted directly within the marketplace interface, where buyers remit payments to sellers. Successful payments trigger several critical updates within the system to reflect the change in product ownership and to maintain the integrity of transaction records. 


\begin{algorithm}[H]
\caption{Payment Process}
\label{algo: payment_process}
\begin{algorithmic}[1]
\STATE $Payment \leftarrow \text{MakePayment}(B_{info},S_{info},p_{id},Pr) $
\IF {$payment == success$} 
    \STATE $DPP \leftarrow \text{UpdateDPP}(B_{info})$
    \STATE $A_{IPFS} \leftarrow \text{StoreInIPFS}(DPP)$
    \STATE $Info \leftarrow {(A_{IPFS}, CPD)}$
    \STATE $T_{id} \leftarrow \text{ChangeOfOwnership}(Info,B_{info} )$
    \STATE $\text{SendToBuyer}(A_{IPFS}, T_{id},CPD)$
\ELSE
     \RETURN "Payment failed."
\ENDIF
\end{algorithmic}
\end{algorithm}

Algorithm~\ref{algo: payment_process} explains the payment process. When the buyer makes the payment, a set of information along with the payment is communicated, such as buyer details (\textit{B\textsubscript{info}}), seller details (\textit{S\textsubscript{info}}), product identifier (\textit {p\textsubscript{id}}), and the price (\textit{Pr}). If the payment is successful, the DPP is updated and stored in IPFS. The reference (\textit{A\textsubscript{IPFS}}) and critical product details (\textit{CPD}) are packaged into an information set (\textit{Info}), which is then recorded on the blockchain with updated ownership. Successful execution of these steps returns \textit{T\textsubscript{id}} and are sent to the buyer along with \textit{A\textsubscript{IPFS}},  and \textit{CPD}. Unsuccessful payment leads to the failure of the payment transaction, and the system state remains unaffected. {Unlike traditional payment workflows, this process incorporates lifecycle continuity by updating the DPP and recording ownership transfer on the blockchain. This ensures that the product’s history is preserved across multiple reuse cycles, which is a key requirement in circular construction practices.}

\section{Framework Evaluation} \label{sec: ch3_evaluation}
All experiments were conducted on a system with 20 GB RAM, an Intel(R) Core(TM) i7-6600U CPU, and running on the Ubuntu 24.04 operating system. Smart contracts were written in \textit{Ink}, which is native to \textit{Substrate} (used to create Polkadot networks), and the IPFS private network is used for decentralised storage. The framework uses MongoDB as the storage for the marketplace details. The backend was developed using \textit{Typescript}, and the frontend was developed using \textit{Vue.js}. The evaluation was performed in two sections: security and performance evaluations. 


\subsection{Security Evaluation} \label{sec: security_evaluation}
{The security analysis presented in this work focuses on application-layer threats specific to the second-hand marketplace, particularly those arising from interactions between sellers, buyers, and the marketplace. These include misrepresentation of product quality, fraudulent claims, and manipulation of transaction data. It is important to distinguish these from infrastructure-level threats inherent to blockchain and decentralised storage systems, such as Sybil attacks, smart contract vulnerabilities, and data availability issues. These threats are typically addressed by the underlying platform through consensus protocols, secure smart contract development practices, and distributed data replication mechanisms \cite{blockchain_attackes}. In this work, we assume a reliable underlying blockchain and IPFS infrastructure and focus on how their properties—immutability, transparency, and verifiability—can be leveraged to mitigate marketplace-specific risks. The use of blockchain references for IPFS-stored DPPs and cross-verification mechanisms ensures that product information cannot be tampered with post-listing, thereby strengthening trust among participants.}

{However, while the framework ensures immutability and traceability of product information after submission, it does not inherently guarantee the correctness of data at the point of entry. To address this, the system adopts an accountability-driven approach, where all submitted information is permanently recorded and can be audited, enabling sellers to be held responsible for any misrepresentation. Incorporating behavioural trust mechanisms such as reputation systems and decentralised identity frameworks are beyond the scope of this work and are considered as a direction for future extension. Such mechanisms can complement the proposed framework by enabling trust-aware interactions based on participant behaviour over time.}
\subsubsection{Malicious Seller}
A malicious seller may attempt to sell a product by hiding its faults or damages from the buyer. The buyer may identify these faults or damages after purchasing the product. By that time, the payment may have been processed, and the seller may not take responsibility for the fault, leaving the buyer at a loss. In the proposed marketplace, we define the behaviours of sellers and the system's response through a sequence of states and transactions to effectively address potential malicious sellers. Initially, a seller may list a product in the state \textit{P\textsubscript{init}}, creating the DPP that includes all relevant product attributes and storing this DPP in the IPFS, with the IPFS reference (\textit{IPFS\textsubscript{ref}}) recorded on the blockchain. This transaction can be described as (\textit{P\textsubscript{init}}, DPP, \textit{P\textsubscript{sell}}) where \textit{P\textsubscript{sell}} represents the state post-listing but pre-transaction confirmation. Upon purchase, if a buyer discovers faults or discrepancies, the state changes to \textit{P\textsubscript{fault}}, triggered by the discovery of the fault described as (\textit{P\textsubscript{sell}}, \textit{discover\textsubscript{fault}}, \textit{P\textsubscript{fault}}). To ascertain the integrity of the DPP, the system engages in a verification process (\textit{P\textsubscript{fault}}, \textit{verify\textsubscript{DPP}}, \textit{P\textsubscript{verify}}) where \textit{P\textsubscript{verify}} is the state in which the DPP is compared with the blockchain-stored \textit{IPFS\textsubscript{ref}}.

The outcome of this verification process is crucial; if the DPP matches the record stored in the blockchain \textit{IPFS\textsubscript{ref}}, it confirms the seller's transparency, and the transaction progresses to a resolution state (\textit{P\textsubscript{verify}}, \textit{match}, \textit{P\textsubscript{resolve}}). Conversely, if there is a mismatch indicating an alteration in the DPP post-listing that is not updated on the blockchain, the system transitions to (\textit{P\textsubscript{verify}}, \textit{mismatch}, \textit{P\textsubscript{fraud}}), highlighting a breach of trust and enabling the implementation of punitive measures against the seller. 

\subsubsection{Malicious Buyer}
A scenario involving a malicious buyer typically revolves around exploiting the returns and refunds processes in a marketplace. A buyer may falsely claim that a received product is defective or not as described, aiming to secure a refund. This type of fraud can lead to significant financial losses for sellers and undermine trust in the marketplace. By applying the same logic for identifying malicious sellers, the framework helps validate or refute the buyer's claims. 

\subsubsection{Malicious Marketplace}
Similar to malicious sellers and buyers, a malicious marketplace scenario may also arise. The marketplace can potentially engage in malicious behaviour, such as manipulating product pricing to its advantage. The scenario and sequence of actions and checks are outlined as follows: Initially, the seller lists a product, sets a price along with the product details, and is then securely logged on the blockchain (\textit{Pr\textsubscript{block}}), ensuring that any subsequent modifications are traceable and auditable. Once the product is added to the blockchain, it is listed on the marketplace with the same price (\textit{Pr\textsubscript{list}}). The seller can lawfully update the price of the product and must ensure that the prices on the blockchain and in the marketplace are the same.


\begin{figure*}[!t]
    \centering
    \begin{subfigure}[b]{0.45\textwidth}
        \centering
        \includegraphics[scale=0.45, trim=3cm 3cm 3cm 3cm, clip]{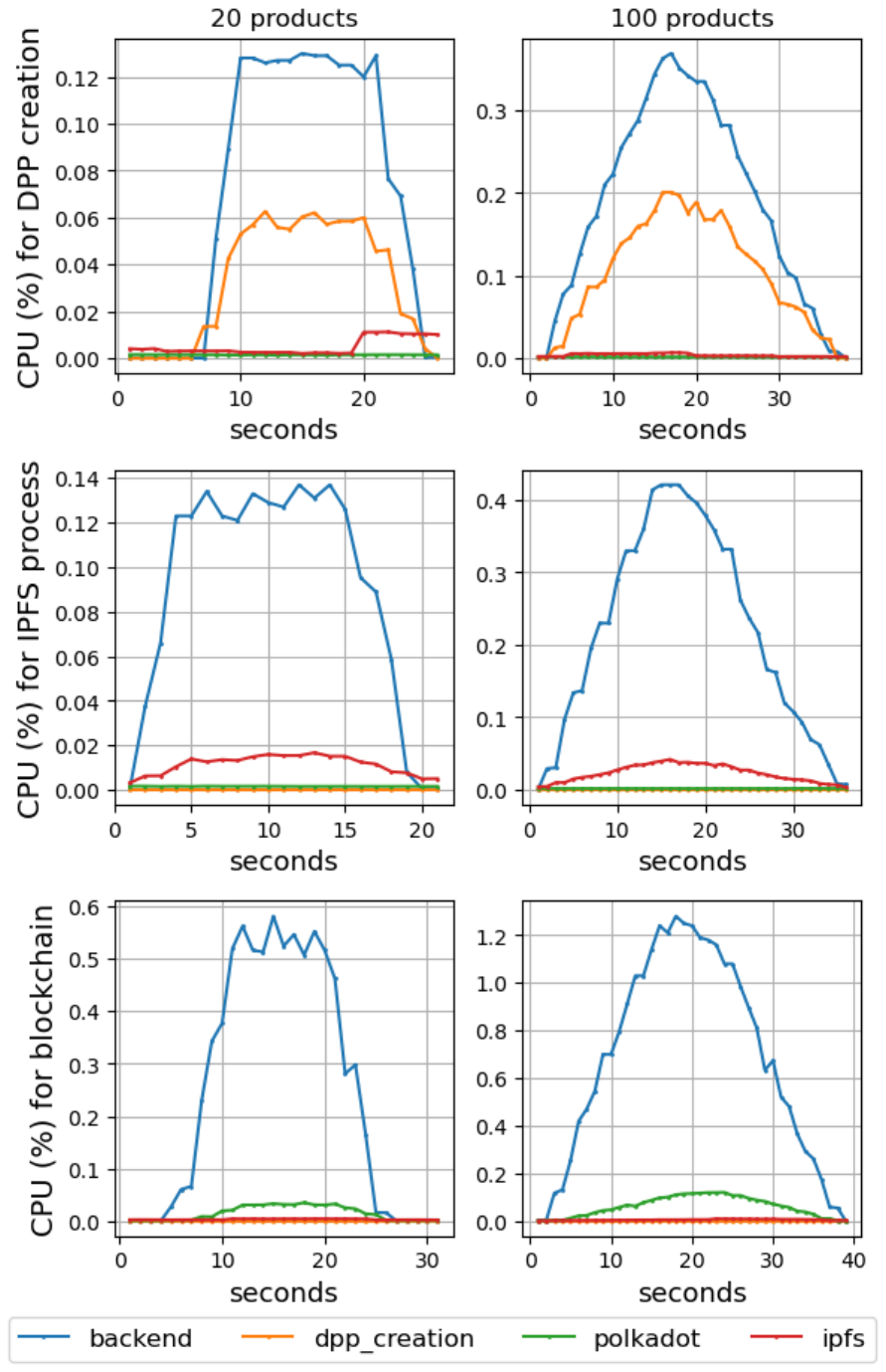}
        \caption{CPU usage Analysis}
        \label{market_cpu}
    \end{subfigure}
    \hfill
    \begin{subfigure}[b]{0.45\textwidth}
        \centering
        \includegraphics[scale=0.45, trim=3cm 3cm 3cm 3cm, clip]{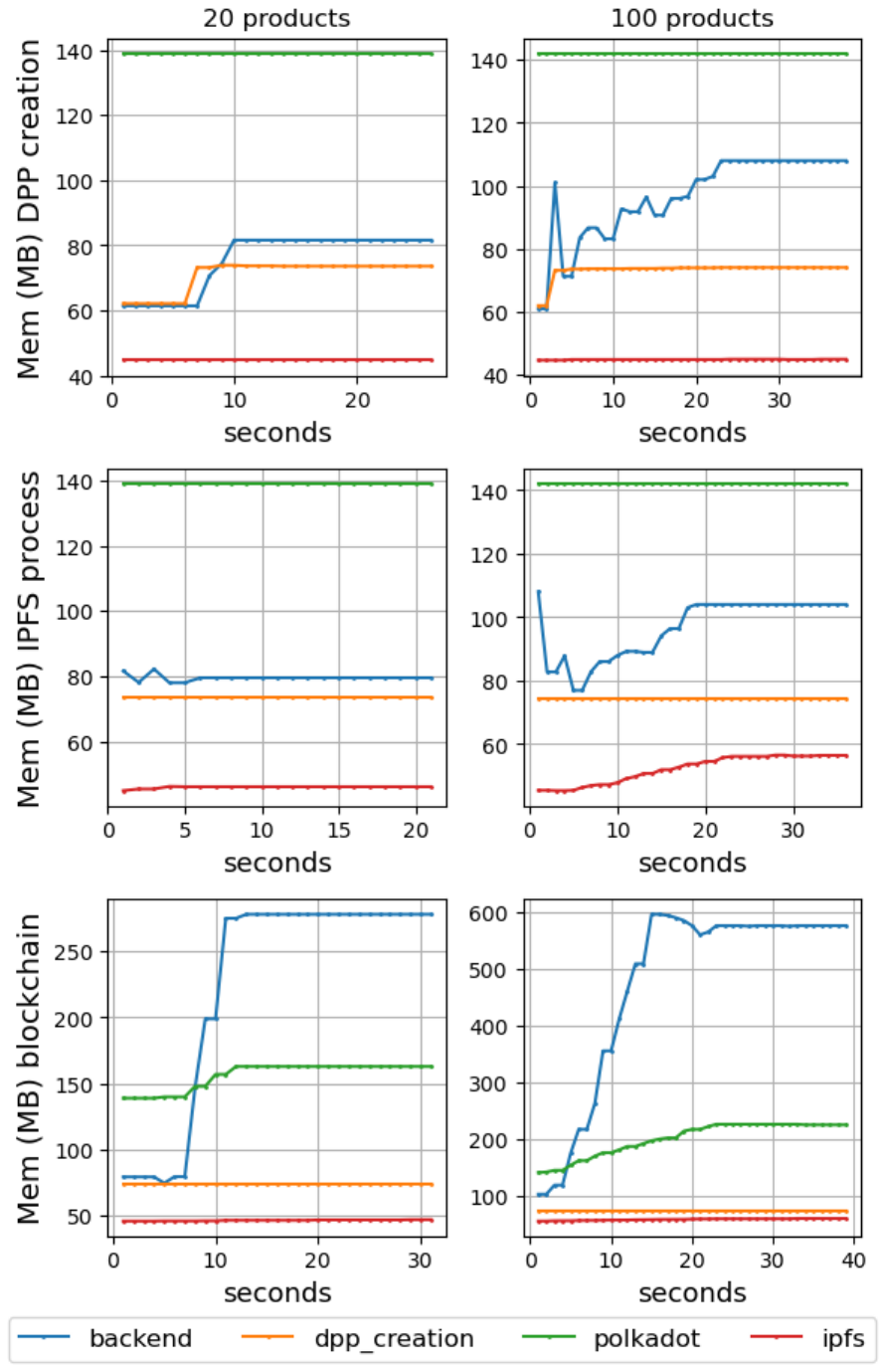}
        \caption{Memory usage Analysis}
        \label{market_memory}
    \end{subfigure}
    \caption{CPU and memory usage across different system components during product processing for 20 and 100 products. The results show that backend services exhibit increased resource utilisation as workload grows, reflecting their central role in coordinating system operations. In contrast, IPFS and blockchain components (Polkadot) demonstrate relatively stable behaviour, indicating predictable performance under increasing load.}
    \label{fig:market_eva_overall}
    \Description{Performance evaluation of the framework with various loads}
\end{figure*}


When a buyer wants to purchase a product listed in the marketplace, the buyer has the opportunity to verify the price recorded on the blockchain. If the price set by the seller and logged on the blockchain matches that of the pricing available on the marketplace (\textit{Pr\textsubscript{block}} =  \textit{Pr\textsubscript{list}}), then the verification process is complete.  Conversely, any discrepancy detected would lead the system to transition to a fraud detection state. This state triggers alerts and potential penalties, reinforcing the system's robustness against marketplace fraud and ensuring protection for both buyers and sellers. 



\subsection{Performance Evaluation}
The individual entities of the framework are containerised using Docker for seamless integration. The performance is evaluated based on its CPU and memory usage for three processes: DPP creation, IPFS process, and blockchain process. The DPP creation process involves generating DPPs from Industry Foundation Classes (IFC) files. This process entails loading an IFC file and extracting relevant details into a DPP, ensuring that all information regarding the materials is properly documented. The IPFS process stores these DPPs in IPFS and returns a reference. The blockchain process stores the IPFS hash and critical product information on a blockchain. The four labels in the figure~\ref{fig:market_eva_overall} represent the four operations that are containerised: \textit{dpp\_creation} refers to the creation of DPP from IFC files, \textit{ipfs} refers to storing of DPP in IPFS, \textit{polkadot} pertains to the recording of information in the blockchain, and \textit{backend} represents the system managing interactions, database operations, and overall coordination between the entities. The CPU and memory usage of the system were analysed while processing 20 and 100 IFC files. {The choice of processing 20 and 100 products is intended to represent two different workload scenarios: a small-scale deployments and a moderate-scale setting to observe system behaviour under increased load. Rather than targeting absolute performance limits, this evaluation focuses on analysing how key system components respond to increasing numbers of products.} Each row in the graph represents the three processes.

Figure~\ref{market_cpu} illustrates the CPU usage of the three processes. In the DPP creation process, the \textit{backend} consistently exhibits the highest CPU across all scenarios, indicating the intensive computation needed to manage the overall process. The \textit{dpp\_creation} process shows moderate CPU utilisation, which increases with the number of products processed. This suggests that generating the DPP has computational demands proportional to the number of products. The \textit{ipfs} and \textit{polkadot} operations display minimal CPU usage during DPP creation, indicating that they are less involved in this stage. 

In the IPFS process, the \textit{backend} has the highest CPU usage, which peaks when handling more products, reflecting its role in coordinating DPP storage on IPFS. The \textit{ipfs} shows low but steady CPU consumption, which indicates its efficiency regardless of the increasing number of products. \textit{dpp\_creation} and \textit{polkadot} exhibit almost negligible CPU usage during this phase. The \textit{backend} also shows the highest CPU utilisation in the third process. The \textit{polkadot} consumes a small but significant amount of CPU, which gradually increases with the number of products, reflecting the computation required to store information on the blockchain. The \textit{ipfs} and \textit{dpp\_creation} exhibit very low CPU usage during blockchain interactions.

Figure~\ref{market_memory} provides insights into the memory requirements for the processes. In DPP creation, the \textit{backend} consistently uses memory, increasing proportionally with the number of products, which is expected because the \textit{backend} handles significant data. The \textit{dpp\_creation} uses relatively stable memory, which remains fairly constant as the number increases. This suggests that the memory for DPP generation is well contained and does not scale significantly. The \textit{ipfs} has almost negligible memory usage, indicating that it is not actively involved in this stage. However, \textit{polkadot} shows a relatively stable but considerable level of memory usage, even though it is not directly involved in creating DPPs. 

In the IPFS process, \textit{polkadot} memory continues to be high and stable, although it is not used during this stage. The \textit{backend} memory usage increases with the number as it communicates with IPFS. The \textit{ipfs} shows a gradual increase in memory consumption, suggesting a slight dependency on the number of products being processed. The \textit{dpp\_creation} has stable memory, although its involvement in this stage is limited. In the blockchain process, the \textit{backend} uses the highest amount of memory, reflecting the complexity of blockchain interactions and transactions. The \textit{polkadot} memory usage is higher than that of the other two processes and gradually increases with the number. The \textit{ipfs} and \textit{dpp\_creation} show minimal memory involvement during this stage. 

{Overall, the \textit{backend} process consistently shows the highest CPU and memory consumption across all three stages, highlighting its central role in coordinating system operations and integrating interactions between components. This trend becomes more pronounced as the number of products increases, indicating that scalability is primarily influenced by backend processing and orchestration. In contrast, components such as \textit{ipfs} and \textit{polkadot} exhibit relatively stable resource usage, suggesting predictable behaviour under increasing workloads. Notably, Polkadot maintains a consistent baseline memory requirement across scenarios, reflecting its steady operational footprint.
These observations suggest that the system can scale effectively with appropriate resource provisioning, particularly by optimising and distributing backend workloads. Given the modular and containerised architecture, horizontal scaling can be achieved by deploying multiple backend instances, IPFS nodes, and blockchain validators, enabling the framework to support larger volumes of products and transactions. While the current evaluation focuses on analysing system behaviour under increasing load, future work will explore large-scale experiments with higher transaction volumes and real-world datasets to further validate scalability in production environments.}

\subsection{Cost Considerations}
The practical adoption of the proposed framework depends not only on its technical capabilities but also on its economic feasibility. This section outlines the key cost components and potential benefits associated with deploying the system in real-world scenarios.

\subsubsection{Cost Components}
{The framework introduces several cost factors. First, infrastructure costs include hosting backend services, maintaining blockchain nodes (e.g., Polkadot/Substrate), and managing databases. Blockchain transaction costs arise from recording product metadata and ownership changes. These costs can be controlled by adopting a permissioned or consortium-based blockchain, reducing reliance on public network fees. Second, decentralised storage costs are associated with maintaining DPP data on IPFS; while IPFS itself is cost-efficient, persistent storage may require pinning services. Finally, integration and operational costs include system deployment, maintenance, and interoperability with existing systems in the industry.}

\subsubsection{Benefits and Value proposition}
{Despite these costs, the framework offers several economic and operational benefits. By enabling trustworthy reuse of construction materials, it can significantly reduce procurement costs and material waste. The immutable record of provenance reduces disputes, fraud, and verification overhead, leading to lower transaction risks. Additionally, improved transparency and compliance tracking can reduce regulatory penalties and auditing costs.}

\subsubsection{Trade-offs and Feasibility}
{The cost-benefit balance depends on the scale of deployment and the value of traded materials. In high-value construction projects, the potential savings from material reuse and reduced risk can outweigh the additional system overhead. Furthermore, consortium-based deployment models can distribute infrastructure costs among stakeholders, improving overall feasibility. While this study provides a qualitative assessment, future work will focus on quantitative cost modelling and pilot deployments to evaluate real-world economic performance.}

\subsection{Limitations of the Proposed Framework}
{While the proposed framework demonstrates the feasibility of integrating blockchain, IPFS, and Digital Product Passports (DPP) within a second-hand marketplace, several limitations must be acknowledged. From a technical perspective, the current performance evaluation is conducted using a prototype deployed in a controlled, containerised environment. The experiments are based on IFC-derived datasets and limited-scale workloads (up to 100 products), which may not fully represent real-world conditions involving high-throughput transactions, large numbers of concurrent users, network latency, and distributed system dynamics. Additionally, the current performance evaluation focuses primarily on CPU and memory utilisation, and does not include other key metrics such as transaction latency, query response time, or end-to-end execution time, which are important for assessing user-level performance. In terms of generalisability, while the framework is designed to be adaptable across domains, the current implementation and evaluation are specific to the built environment. Further validation in other domains and with real-world datasets is required to confirm broader applicability.}

{From a scientific perspective, the proposed system does not include formal verification or model-checking of smart contracts and workflow logic. As a result, properties such as safety, liveness, and correctness under all possible system states are not formally guaranteed. Additionally, while the framework ensures data integrity and accountability, it does not guarantee the correctness of data at the point of entry, as discussed in Section~\ref{sec: security_evaluation}. Addressing these limitations through large-scale deployment, real-world validation, and formal verification techniques represents an important direction for future research. Additionally, the current study does not include a quantitative benchmarking against existing centralised or decentralised marketplace systems. While a qualitative comparison is provided in Section~\ref{sec: marketplace_literature}, future work will focus on conducting comparative evaluations to assess improvements in efficiency, scalability, and trust under real-world conditions. The governance structure of the consortium, including mechanisms for rule updates, dispute resolution, and incentive or commission models, is not explicitly defined in the current work. Ongoing research explores extensions of the framework, including auction-based mechanisms for dynamic pricing\cite{auction_stanly} and tokenised incentive models (e.g., carbon credits)\cite{carbon_credit} to support reward and commission structures. These directions aim to enhance the economic and governance aspects of the marketplace.}

\section{Conclusion and Future Work-} \label{sec: ch3_conclusion}
In conclusion, the proposed blockchain-enabled marketplace enhances operational efficiency and trust in industrial reuse markets. By improving traceability and reducing reliance on intermediaries, the framework addresses key challenges in circular supply chains and contributes to sustainable and automated industrial processes.
By leveraging blockchain technology, the platform ensures transparency, traceability, and accountability in transactions, addressing key challenges such as verifying product quality and origin. The prototype aims to facilitate a reliable mechanism for listing and verifying reusable materials in the marketplace. 
The effectiveness of the marketplace depends significantly on the balanced interaction between supply and demand, and this imbalance can affect the marketplace's viability and user satisfaction. 

The evaluation results demonstrate that the proposed framework maintains stable and predictable behaviour as the number of products increases, with resource utilisation primarily driven by backend coordination, while IPFS and blockchain components remain consistent. This indicates that the system can scale effectively with appropriate resource provisioning. Furthermore, the security analysis shows that the framework can successfully detect and mitigate marketplace-specific threats, such as data manipulation and misrepresentation, through verifiable DPP records and blockchain-based traceability, thereby reinforcing trust among participants.

{While this work focuses on the built environment, the proposed framework is designed to be adaptable across domains where product provenance and reuse are critical. The core components—blockchain-based traceability, IPFS-backed DPP, and marketplace mechanisms—are domain-agnostic. Domain-specific adaptations can be achieved by replacing the CCPO with relevant ontologies and evaluation criteria. For instance, in electronics, DPPs may include component-level diagnostics and refurbishment history; in apparel, fibre composition and sustainability certifications; and in battery ecosystems, lifecycle indicators such as charge cycles and degradation levels. This demonstrates the scalability of the framework as a generalisable architecture for trustworthy second-hand marketplaces.}

{Future work will focus on extending the framework to support additional functionalities required for real-world deployment. This includes incorporating compliance checks to ensure that listed products meet regulatory standards, integrating dynamic pricing mechanisms based on supply, demand, and product condition, and exploring behavioural trust mechanisms such as reputation systems to enhance participant reliability. The future research will investigate approaches to address the data oracle problem, including the use of trusted data sources, certification processes and verification, to improve the alignment between physical assets and their digital representations. Additionally, large-scale evaluations using real-world datasets and high-throughput scenarios will be conducted to further validate system performance and scalability.}



\begin{acks}
This work is supported in part by the Engineering and Physical Sciences Research Council "Digital Economy" programme: EP/V042521/1 and EP/V042017/1. We would like to express our gratitude to BOHM (Dara Khera, Liam Tootill, Sukshma Paranjpe and Allen Mak) who provided valuable insights for this work. 
\end{acks}

\bibliographystyle{ACM-Reference-Format}
\bibliography{bibliography}


\begin{thebibliography}{33}


\ifx \showCODEN    \undefined \def \showCODEN     #1{\unskip}     \fi
\ifx \showISBNx    \undefined \def \showISBNx     #1{\unskip}     \fi
\ifx \showISBNxiii \undefined \def \showISBNxiii  #1{\unskip}     \fi
\ifx \showISSN     \undefined \def \showISSN      #1{\unskip}     \fi
\ifx \showLCCN     \undefined \def \showLCCN      #1{\unskip}     \fi
\ifx \shownote     \undefined \def \shownote      #1{#1}          \fi
\ifx \showarticletitle \undefined \def \showarticletitle #1{#1}   \fi
\ifx \showURL      \undefined \def \showURL       {\relax}        \fi
\providecommand\bibfield[2]{#2}
\providecommand\bibinfo[2]{#2}
\providecommand\natexlab[1]{#1}
\providecommand\showeprint[2][]{arXiv:#2}

\bibitem[Adu-Duodu et~al\mbox{.}(2025a)]%
        {carbon_credit}
\bibfield{author}{\bibinfo{person}{Kwabena Adu-Duodu}, \bibinfo{person}{Dewa Gde~Adi Murthi~Udayana}, \bibinfo{person}{Stanly Wilson}, \bibinfo{person}{Yinhao Li}, \bibinfo{person}{Rajiv Ranjan}, {and} \bibinfo{person}{Ellis Solaiman}.} \bibinfo{year}{2025}\natexlab{a}.
\newblock \bibinfo{title}{Leveraging Tokenised Carbon Credits to Drive Circularity in Supply Chains}.
\newblock \bibinfo{howpublished}{\url{https://ssrn.com/abstract=5913126 or http://dx.doi.org/10.2139/ssrn.5913126}}.
\newblock
\newblock
\shownote{Accessed April 20, 2026.}.


\bibitem[Adu-Duodu et~al\mbox{.}(2025b)]%
        {ccpo}
\bibfield{author}{\bibinfo{person}{Kwabena Adu-Duodu}, \bibinfo{person}{Stanly Wilson}, \bibinfo{person}{Yinhao Li}, \bibinfo{person}{Aanuoluwapo Oladimeji}, \bibinfo{person}{Talea Huraysi}, \bibinfo{person}{Masoud Barati}, \bibinfo{person}{Charith Perera}, \bibinfo{person}{Ellis Solaiman}, \bibinfo{person}{Omer Rana}, \bibinfo{person}{Rajiv Ranjan}, {and} \bibinfo{person}{Tejal Shah}.} \bibinfo{year}{2025}\natexlab{b}.
\newblock \showarticletitle{A Circular Construction Product Ontology for End-of-Life Decision-Making}. In \bibinfo{booktitle}{\emph{The 40th ACM/SIGAPP Symposium On Applied Computing (SAC)}}. \bibinfo{publisher}{ACM}.
\newblock
\href{https://doi.org/10.48550/arXiv.2503.13708}{doi:\nolinkurl{10.48550/arXiv.2503.13708}}


\bibitem[Ahmadjee et~al\mbox{.}(2025)]%
        {blockchain_attackes}
\bibfield{author}{\bibinfo{person}{Sabreen Ahmadjee}, \bibinfo{person}{Carlos Mera-G{\'o}mez}, \bibinfo{person}{Rami Bahsoon}, {and} \bibinfo{person}{Rajkumar Buyya}.} \bibinfo{year}{2025}\natexlab{}.
\newblock \showarticletitle{Security Architectural Approaches and Risk Assessment Methods for Blockchain Systems: A Review and Future Directions}.
\newblock \bibinfo{journal}{\emph{Distributed Ledger Technologies: Research and Practice}} \bibinfo{volume}{5}, \bibinfo{number}{1} (\bibinfo{year}{2025}), \bibinfo{pages}{1--21}.
\newblock
\href{https://doi.org/10.1145/3721140}{doi:\nolinkurl{10.1145/3721140}}


\bibitem[Banerjee et~al\mbox{.}(2023)]%
        {acmdlt_fair_exchange}
\bibfield{author}{\bibinfo{person}{Prabal Banerjee}, \bibinfo{person}{Dushyant Behl}, \bibinfo{person}{Palanivel Kodeswaran}, \bibinfo{person}{Chaitanya Kumar}, \bibinfo{person}{Sushmita Ruj}, \bibinfo{person}{Sayandeep Sen}, {and} \bibinfo{person}{Dhinakaran Vinayagamurthy}.} \bibinfo{year}{2023}\natexlab{}.
\newblock \showarticletitle{Accelerated Verifiable Fair Digital Exchange}.
\newblock \bibinfo{journal}{\emph{Distributed Ledger Technologies: Research and Practice}} \bibinfo{volume}{2}, \bibinfo{number}{3} (\bibinfo{year}{2023}).
\newblock
\href{https://doi.org/10.1145/3596448}{doi:\nolinkurl{10.1145/3596448}}


\bibitem[Banerjee and Ruj(2018)]%
        {marketplaces_data_arxiv}
\bibfield{author}{\bibinfo{person}{Prabal Banerjee} {and} \bibinfo{person}{Sushmita Ruj}.} \bibinfo{year}{2018}\natexlab{}.
\newblock \bibinfo{title}{Blockchain enabled data marketplace--design and challenges}.
\newblock
\href{https://doi.org/10.48550/arXiv.1811.11462}{doi:\nolinkurl{10.48550/arXiv.1811.11462}}
\showeprint[arxiv]{1811.11462}~[cs.CR]


\bibitem[Caldera et~al\mbox{.}(2020)]%
        {barriers_enablers_sustainability}
\bibfield{author}{\bibinfo{person}{Savindi Caldera}, \bibinfo{person}{Tim Ryley}, {and} \bibinfo{person}{Nikita Zatyko}.} \bibinfo{year}{2020}\natexlab{}.
\newblock \showarticletitle{Enablers and Barriers for Creating a Marketplace for Construction and Demolition Waste: A Systematic Literature Review}.
\newblock \bibinfo{journal}{\emph{Sustainability}} \bibinfo{volume}{12}, \bibinfo{number}{23} (\bibinfo{year}{2020}).
\newblock
\href{https://doi.org/10.3390/su12239931}{doi:\nolinkurl{10.3390/su12239931}}


\bibitem[Cano et~al\mbox{.}(2023)]%
        {trust_e-marketplaces}
\bibfield{author}{\bibinfo{person}{Jose~Alejandro Cano}, \bibinfo{person}{Abraham~Allec Londoño-Pineda}, \bibinfo{person}{Emiro~Antonio Campo}, {and} \bibinfo{person}{Sergio~Augusto Fernández}.} \bibinfo{year}{2023}\natexlab{}.
\newblock \showarticletitle{Sustainable business models of e-marketplaces: An analysis from the consumer perspective}.
\newblock \bibinfo{journal}{\emph{Journal of Open Innovation: Technology, Market, and Complexity}} \bibinfo{volume}{9}, \bibinfo{number}{3} (\bibinfo{year}{2023}), \bibinfo{pages}{100121}.
\newblock
\href{https://doi.org/10.1016/j.joitmc.2023.100121}{doi:\nolinkurl{10.1016/j.joitmc.2023.100121}}


\bibitem[Dixit et~al\mbox{.}(2023)]%
        {fast_data}
\bibfield{author}{\bibinfo{person}{Akanksha Dixit}, \bibinfo{person}{Arjun Singh}, \bibinfo{person}{Yogachandran Rahulamathavan}, {and} \bibinfo{person}{Muttukrishnan Rajarajan}.} \bibinfo{year}{2023}\natexlab{}.
\newblock \showarticletitle{FAST DATA: A Fair, Secure, and Trusted Decentralized IIoT Data Marketplace Enabled by Blockchain}.
\newblock \bibinfo{journal}{\emph{IEEE Internet of Things Journal}} \bibinfo{volume}{10}, \bibinfo{number}{4} (\bibinfo{year}{2023}), \bibinfo{pages}{2934--2944}.
\newblock
\href{https://doi.org/10.1109/JIOT.2021.3120640}{doi:\nolinkurl{10.1109/JIOT.2021.3120640}}


\bibitem[Fang and Bai(2024)]%
        {refurbished}
\bibfield{author}{\bibinfo{person}{Zhou Fang} {and} \bibinfo{person}{Yang Bai}.} \bibinfo{year}{2024}\natexlab{}.
\newblock \showarticletitle{A comparative analysis of channel strategies for refurbished products: The influence of blockchain, brand spillover on co-opetitive supply chain}.
\newblock \bibinfo{journal}{\emph{Managerial and Decision Economics}} \bibinfo{volume}{45}, \bibinfo{number}{4} (\bibinfo{year}{2024}), \bibinfo{pages}{2042--2058}.
\newblock
\href{https://doi.org/10.1002/mde.4118}{doi:\nolinkurl{10.1002/mde.4118}}


\bibitem[González et~al\mbox{.}(2023)]%
        {iot_data_BIDM}
\bibfield{author}{\bibinfo{person}{Víctor González}, \bibinfo{person}{Luis Sánchez}, \bibinfo{person}{Jorge Lanza}, \bibinfo{person}{Juan~Ramón Santana}, \bibinfo{person}{Pablo Sotres}, {and} \bibinfo{person}{Alberto~E. García}.} \bibinfo{year}{2023}\natexlab{}.
\newblock \showarticletitle{On the use of Blockchain to enable a highly scalable Internet of Things Data Marketplace}.
\newblock \bibinfo{journal}{\emph{Internet of Things}}  \bibinfo{volume}{22} (\bibinfo{year}{2023}), \bibinfo{pages}{100722}.
\newblock
\href{https://doi.org/10.1016/j.iot.2023.100722}{doi:\nolinkurl{10.1016/j.iot.2023.100722}}


\bibitem[Haq and Alam(2023)]%
        {cloth_reuse}
\bibfield{author}{\bibinfo{person}{Upama~Nasrin Haq} {and} \bibinfo{person}{S.M.~Rakifull Alam}.} \bibinfo{year}{2023}\natexlab{}.
\newblock \showarticletitle{Implementing circular economy principles in the apparel production process: Reusing pre-consumer waste for sustainability of environment and economy}.
\newblock \bibinfo{journal}{\emph{Cleaner Waste Systems}}  \bibinfo{volume}{6} (\bibinfo{year}{2023}), \bibinfo{pages}{100108}.
\newblock
\href{https://doi.org/10.1016/j.clwas.2023.100108}{doi:\nolinkurl{10.1016/j.clwas.2023.100108}}


\bibitem[Helander and Ljunggren(2023)]%
        {batteries_reuse}
\bibfield{author}{\bibinfo{person}{Harald Helander} {and} \bibinfo{person}{Maria Ljunggren}.} \bibinfo{year}{2023}\natexlab{}.
\newblock \showarticletitle{Battery as a service: Analysing multiple reuse and recycling loops}.
\newblock \bibinfo{journal}{\emph{Resources, Conservation and Recycling}}  \bibinfo{volume}{197} (\bibinfo{year}{2023}), \bibinfo{pages}{107091}.
\newblock
\href{https://doi.org/10.1016/j.resconrec.2023.107091}{doi:\nolinkurl{10.1016/j.resconrec.2023.107091}}


\bibitem[Incorvaja et~al\mbox{.}(2022)]%
        {Omer_circular}
\bibfield{author}{\bibinfo{person}{Dan Incorvaja}, \bibinfo{person}{Yasin Celik}, \bibinfo{person}{Ioan Petri}, {and} \bibinfo{person}{Omer Rana}.} \bibinfo{year}{2022}\natexlab{}.
\newblock \showarticletitle{Circular Economy and Construction Supply Chains}. In \bibinfo{booktitle}{\emph{2022 IEEE/ACM International Conference on Big Data Computing, Applications and Technologies (BDCAT)}}. \bibinfo{publisher}{IEEE/ACM}, \bibinfo{pages}{92--99}.
\newblock
\href{https://doi.org/10.1109/BDCAT56447.2022.00019}{doi:\nolinkurl{10.1109/BDCAT56447.2022.00019}}


\bibitem[Knoth et~al\mbox{.}(2022)]%
        {barriers_norway}
\bibfield{author}{\bibinfo{person}{Katrin Knoth}, \bibinfo{person}{Selamawit~Mamo Fufa}, {and} \bibinfo{person}{Erlend Seilskjær}.} \bibinfo{year}{2022}\natexlab{}.
\newblock \showarticletitle{Barriers, success factors, and perspectives for the reuse of construction products in Norway}.
\newblock \bibinfo{journal}{\emph{Journal of Cleaner Production}}  \bibinfo{volume}{337} (\bibinfo{year}{2022}), \bibinfo{pages}{130494}.
\newblock
\href{https://doi.org/10.1016/j.jclepro.2022.130494}{doi:\nolinkurl{10.1016/j.jclepro.2022.130494}}


\bibitem[Lamela et~al\mbox{.}(2022)]%
        {IEEE_access_developing_trustworthy}
\bibfield{author}{\bibinfo{person}{Manuel~Pereira Lamela}, \bibinfo{person}{Jesús Rodríguez-Molina}, \bibinfo{person}{Margarita Martínez-Núñez}, {and} \bibinfo{person}{Juan Garbajosa}.} \bibinfo{year}{2022}\natexlab{}.
\newblock \showarticletitle{A Blockchain-Based Decentralized Marketplace for Trustworthy Trade in Developing Countries}.
\newblock \bibinfo{journal}{\emph{IEEE Access}}  \bibinfo{volume}{10} (\bibinfo{year}{2022}), \bibinfo{pages}{79100--79123}.
\newblock
\href{https://doi.org/10.1109/ACCESS.2022.3194511}{doi:\nolinkurl{10.1109/ACCESS.2022.3194511}}


\bibitem[Li et~al\mbox{.}(2023)]%
        {TEM_luxuary}
\bibfield{author}{\bibinfo{person}{Guangming Li}, \bibinfo{person}{Zhi-Ping Fan}, {and} \bibinfo{person}{Xue-Yan Wu}.} \bibinfo{year}{2023}\natexlab{}.
\newblock \showarticletitle{The Choice Strategy of Authentication Technology for Luxury E-Commerce Platforms in the Blockchain Era}.
\newblock \bibinfo{journal}{\emph{IEEE Transactions on Engineering Management}} \bibinfo{volume}{70}, \bibinfo{number}{3} (\bibinfo{year}{2023}), \bibinfo{pages}{1239--1252}.
\newblock
\href{https://doi.org/10.1109/TEM.2021.3076606}{doi:\nolinkurl{10.1109/TEM.2021.3076606}}


\bibitem[Liu et~al\mbox{.}(2024)]%
        {reseller_market}
\bibfield{author}{\bibinfo{person}{Yun Liu}, \bibinfo{person}{Deqing Ma}, {and} \bibinfo{person}{Jinsong Hu}.} \bibinfo{year}{2024}\natexlab{}.
\newblock \showarticletitle{Reseller or Market? E-platform strategy under blockchain-enabled recycling}.
\newblock \bibinfo{journal}{\emph{Expert Systems with Applications}}  \bibinfo{volume}{243} (\bibinfo{year}{2024}), \bibinfo{pages}{122811}.
\newblock
\href{https://doi.org/10.1016/j.eswa.2023.122811}{doi:\nolinkurl{10.1016/j.eswa.2023.122811}}


\bibitem[Mahmoud et~al\mbox{.}(2024)]%
        {acmdlt_iot_reputaion}
\bibfield{author}{\bibinfo{person}{Haitham Mahmoud}, \bibinfo{person}{Junaid Arshad}, {and} \bibinfo{person}{Adel Aneiba}.} \bibinfo{year}{2024}\natexlab{}.
\newblock \showarticletitle{A Systematic Review of Blockchain-Based Privacy-Preserving Reputation Systems for IoT Applications}.
\newblock \bibinfo{journal}{\emph{Distributed Ledger Technologies: Research and Practice}} \bibinfo{volume}{3}, \bibinfo{number}{4} (\bibinfo{year}{2024}).
\newblock
\href{https://doi.org/10.1145/3674156}{doi:\nolinkurl{10.1145/3674156}}


\bibitem[Mavrogiorgou et~al\mbox{.}(2023)]%
        {Fame}
\bibfield{author}{\bibinfo{person}{Argyro Mavrogiorgou}, \bibinfo{person}{Athanasios Kiourtis}, \bibinfo{person}{Georgios Makridis}, \bibinfo{person}{Dimitrios Kotios}, \bibinfo{person}{Vasileios Koukos}, \bibinfo{person}{Dimosthenis Kyriazis}, \bibinfo{person}{John Soldatos}, \bibinfo{person}{Georgios Fatouros}, \bibinfo{person}{Dimitrios Drakoulis}, \bibinfo{person}{Pedro Maló}, \bibinfo{person}{Martin Serrano}, \bibinfo{person}{Mauro Isaja}, \bibinfo{person}{Raquel Lazcano}, \bibinfo{person}{Juan~Manuel Vera}, \bibinfo{person}{Fabiana Fournier}, \bibinfo{person}{Lior Limonad}, \bibinfo{person}{Konstantinos Perakis}, \bibinfo{person}{Dimitrios Miltiadou}, \bibinfo{person}{Pavlos Kranas}, \bibinfo{person}{Alper Sen}, \bibinfo{person}{Irene Zattarin}, {and} \bibinfo{person}{Ernesto Troiano}.} \bibinfo{year}{2023}\natexlab{}.
\newblock \showarticletitle{FAME: Federated Decentralized Trusted Data Marketplace for Embedded Finance}. In \bibinfo{booktitle}{\emph{2023 International Conference on Smart Applications, Communications and Networking (SmartNets)}}. \bibinfo{publisher}{IEEE}, \bibinfo{pages}{1--6}.
\newblock
\href{https://doi.org/10.1109/SmartNets58706.2023.10215814}{doi:\nolinkurl{10.1109/SmartNets58706.2023.10215814}}


\bibitem[Michalopoulos et~al\mbox{.}(2023)]%
        {marketplaces_iotdata}
\bibfield{author}{\bibinfo{person}{Panagiotis Michalopoulos}, \bibinfo{person}{Srisht~Fateh Singh}, {and} \bibinfo{person}{Andreas Veneris}.} \bibinfo{year}{2023}\natexlab{}.
\newblock \showarticletitle{Inducing trust in Blockchain-enabled IoT marketplaces through reputation and dispute resolution}. In \bibinfo{booktitle}{\emph{2023 IEEE International Conference on Metaverse Computing, Networking and Applications (MetaCom)}}. \bibinfo{publisher}{IEEE}, \bibinfo{pages}{398--402}.
\newblock


\bibitem[Miller(2021)]%
        {bbc_construction_waste}
\bibfield{author}{\bibinfo{person}{Norman Miller}.} \bibinfo{year}{16/12/2021}\natexlab{}.
\newblock \bibinfo{title}{Extracting materials is wreaking havoc on the planet. Could the world's growing mounds of waste hold the key to sustainable construction?}
\newblock \bibinfo{howpublished}{\url{https://www.bbc.com/future/article/20211215-the-buildings-made-from-rubbish}}.
\newblock
\newblock
\shownote{Accessed March 30, 2026.}.


\bibitem[Sarpatwar et~al\mbox{.}(2019)]%
        {AI_marketplace}
\bibfield{author}{\bibinfo{person}{Kanthi Sarpatwar}, \bibinfo{person}{Venkata~Sitaramagiridharganesh Ganapavarapu}, \bibinfo{person}{Karthikeyan Shanmugam}, \bibinfo{person}{Akond Rahman}, {and} \bibinfo{person}{Roman Vaculin}.} \bibinfo{year}{2019}\natexlab{}.
\newblock \showarticletitle{Blockchain Enabled AI Marketplace: The Price You Pay for Trust}. In \bibinfo{booktitle}{\emph{2019 IEEE/CVF Conference on Computer Vision and Pattern Recognition Workshops (CVPRW)}}. \bibinfo{publisher}{IEEE}, \bibinfo{pages}{2857--2866}.
\newblock
\href{https://doi.org/10.1109/CVPRW.2019.00345}{doi:\nolinkurl{10.1109/CVPRW.2019.00345}}


\bibitem[Schepler et~al\mbox{.}(2024)]%
        {consumer_electronics_reuse}
\bibfield{author}{\bibinfo{person}{Xavier Schepler}, \bibinfo{person}{Nabil Absi}, {and} \bibinfo{person}{Antoine Jeanjean}.} \bibinfo{year}{2024}\natexlab{}.
\newblock \showarticletitle{Refurbishment and remanufacturing planning model for pre-owned consumer electronics}.
\newblock \bibinfo{journal}{\emph{International Journal of Production Research}} \bibinfo{volume}{62}, \bibinfo{number}{7} (\bibinfo{year}{2024}), \bibinfo{pages}{2499--2521}.
\newblock
\href{https://doi.org/10.1080/00207543.2023.2218942}{doi:\nolinkurl{10.1080/00207543.2023.2218942}}


\bibitem[Shi et~al\mbox{.}(2021)]%
        {awesome_conference}
\bibfield{author}{\bibinfo{person}{Zeshun Shi}, \bibinfo{person}{Siamak Farshidi}, \bibinfo{person}{Huan Zhou}, {and} \bibinfo{person}{Zhiming Zhao}.} \bibinfo{year}{2021}\natexlab{}.
\newblock \showarticletitle{An Auction and Witness Enhanced Trustworthy SLA Model for Decentralized Cloud Marketplaces}. In \bibinfo{booktitle}{\emph{Proceedings of the Conference on Information Technology for Social Good}} (Roma, Italy) \emph{(\bibinfo{series}{GoodIT '21})}. \bibinfo{publisher}{ACM}, \bibinfo{pages}{109–114}.
\newblock
\showISBNx{9781450384780}
\href{https://doi.org/10.1145/3462203.3475876}{doi:\nolinkurl{10.1145/3462203.3475876}}


\bibitem[Tan and Saraniemi(2023)]%
        {blockchain_exchange}
\bibfield{author}{\bibinfo{person}{Teck~Ming Tan} {and} \bibinfo{person}{Saila Saraniemi}.} \bibinfo{year}{2023}\natexlab{}.
\newblock \showarticletitle{Trust in blockchain-enabled exchanges: Future directions in blockchain marketing}.
\newblock \bibinfo{journal}{\emph{Journal of the Academy of Marketing Science}} \bibinfo{volume}{51}, \bibinfo{number}{4} (\bibinfo{year}{2023}), \bibinfo{pages}{914--939}.
\newblock
\href{https://doi.org/10.1007/s11747-022-00889-0}{doi:\nolinkurl{10.1007/s11747-022-00889-0}}


\bibitem[Treleaven et~al\mbox{.}(2021)]%
        {real_estate_marketplace}
\bibfield{author}{\bibinfo{person}{Philip Treleaven}, \bibinfo{person}{Jeremy Barnett}, \bibinfo{person}{Andrew Knight}, {and} \bibinfo{person}{Will Serrano}.} \bibinfo{year}{2021}\natexlab{}.
\newblock \showarticletitle{Real estate data marketplace}.
\newblock \bibinfo{journal}{\emph{AI and Ethics}}  \bibinfo{volume}{1} (\bibinfo{year}{2021}), \bibinfo{pages}{445--462}.
\newblock
\href{https://doi.org/10.1007/s43681-021-00053-4}{doi:\nolinkurl{10.1007/s43681-021-00053-4}}


\bibitem[Wang et~al\mbox{.}(2019)]%
        {artchain}
\bibfield{author}{\bibinfo{person}{Ziyuan Wang}, \bibinfo{person}{Lin Yang}, \bibinfo{person}{Qin Wang}, \bibinfo{person}{Donghai Liu}, \bibinfo{person}{Zhiyu Xu}, {and} \bibinfo{person}{Shigang Liu}.} \bibinfo{year}{2019}\natexlab{}.
\newblock \showarticletitle{ArtChain: Blockchain-Enabled Platform for Art Marketplace}. In \bibinfo{booktitle}{\emph{2019 IEEE International Conference on Blockchain (Blockchain)}}. \bibinfo{publisher}{IEEE}, \bibinfo{pages}{447--454}.
\newblock
\href{https://doi.org/10.1109/Blockchain.2019.00068}{doi:\nolinkurl{10.1109/Blockchain.2019.00068}}


\bibitem[Wilson et~al\mbox{.}(2024)]%
        {stanly_future_internet}
\bibfield{author}{\bibinfo{person}{Stanly Wilson}, \bibinfo{person}{Kwabena Adu-Duodu}, \bibinfo{person}{Yinhao Li}, \bibinfo{person}{Ringo Sham}, \bibinfo{person}{Mohammed Almubarak}, \bibinfo{person}{Yingli Wang}, \bibinfo{person}{Ellis Solaiman}, \bibinfo{person}{Charith Perera}, \bibinfo{person}{Rajiv Ranjan}, {and} \bibinfo{person}{Omer Rana}.} \bibinfo{year}{2024}\natexlab{}.
\newblock \showarticletitle{Blockchain-Enabled Provenance Tracking for Sustainable Material Reuse in Construction Supply Chains}.
\newblock \bibinfo{journal}{\emph{Future Internet}} \bibinfo{volume}{16}, \bibinfo{number}{4} (\bibinfo{year}{2024}).
\newblock
\href{https://doi.org/10.3390/fi16040135}{doi:\nolinkurl{10.3390/fi16040135}}


\bibitem[Wilson et~al\mbox{.}(2023)]%
        {short_paper}
\bibfield{author}{\bibinfo{person}{Stanly Wilson}, \bibinfo{person}{Kwabena Adu-Duodu}, \bibinfo{person}{Yinhao Li}, \bibinfo{person}{Ringo Sham}, \bibinfo{person}{Yingli Wang}, \bibinfo{person}{Ellis Solaiman}, \bibinfo{person}{Charith Perera}, \bibinfo{person}{Rajiv Ranjan}, {and} \bibinfo{person}{Omer Rana}.} \bibinfo{year}{2023}\natexlab{}.
\newblock \showarticletitle{Tracking Material Reuse across Construction Supply Chains}. In \bibinfo{booktitle}{\emph{2023 IEEE 19th International Conference on e-Science (e-Science)}}. \bibinfo{publisher}{IEEE}, \bibinfo{pages}{1--4}.
\newblock
\href{https://doi.org/10.1109/e-Science58273.2023.10254935}{doi:\nolinkurl{10.1109/e-Science58273.2023.10254935}}


\bibitem[Wilson et~al\mbox{.}(2025)]%
        {auction_stanly}
\bibfield{author}{\bibinfo{person}{Stanly Wilson}, \bibinfo{person}{Tanapon Suwankesawong}, \bibinfo{person}{Kwabena Adu-Duodu}, \bibinfo{person}{Yinhao Li}, \bibinfo{person}{Rajiv Ranjan}, \bibinfo{person}{Omer Rana}, {and} \bibinfo{person}{Ellis Solaiman}.} \bibinfo{year}{2025}\natexlab{}.
\newblock \showarticletitle{Verifiable Auction Mechanisms for Material Reuse in the Built Environment}. In \bibinfo{booktitle}{\emph{2025 7th International Conference on Blockchain Computing and Applications (BCCA)}}. \bibinfo{pages}{895--897}.
\newblock
\href{https://doi.org/10.1109/BCCA66705.2025.11229665}{doi:\nolinkurl{10.1109/BCCA66705.2025.11229665}}


\bibitem[Yevu et~al\mbox{.}(2021)]%
        {digital_construction}
\bibfield{author}{\bibinfo{person}{Sitsofe~Kwame Yevu}, \bibinfo{person}{Ann~T.W. Yu}, {and} \bibinfo{person}{Amos Darko}.} \bibinfo{year}{2021}\natexlab{}.
\newblock \showarticletitle{Digitalization of construction supply chain and procurement in the built environment: Emerging technologies and opportunities for sustainable processes}.
\newblock \bibinfo{journal}{\emph{Journal of Cleaner Production}}  \bibinfo{volume}{322} (\bibinfo{year}{2021}), \bibinfo{pages}{129093}.
\newblock
\href{https://doi.org/10.1016/j.jclepro.2021.129093}{doi:\nolinkurl{10.1016/j.jclepro.2021.129093}}


\bibitem[Zhang and Seuring(2024)]%
        {DPP_survey}
\bibfield{author}{\bibinfo{person}{Abraham Zhang} {and} \bibinfo{person}{Stefan Seuring}.} \bibinfo{year}{2024}\natexlab{}.
\newblock \showarticletitle{Digital product passport for sustainable and circular supply chain management: a structured review of use cases}.
\newblock \bibinfo{journal}{\emph{International Journal of Logistics Research and Applications}} (\bibinfo{year}{2024}), \bibinfo{pages}{1--28}.
\newblock
\href{https://doi.org/10.1080/13675567.2024.2374256}{doi:\nolinkurl{10.1080/13675567.2024.2374256}}


\bibitem[Zhang et~al\mbox{.}(2024)]%
        {payment_polkadot}
\bibfield{author}{\bibinfo{person}{Peiyun Zhang}, \bibinfo{person}{Xiaoqi Hua}, {and} \bibinfo{person}{Haibin Zhu}.} \bibinfo{year}{2024}\natexlab{}.
\newblock \showarticletitle{Cross-Chain Digital Asset System for Secure Trading and Payment}.
\newblock \bibinfo{journal}{\emph{IEEE Transactions on Computational Social Systems}} \bibinfo{volume}{11}, \bibinfo{number}{2} (\bibinfo{year}{2024}), \bibinfo{pages}{1654--1666}.
\newblock
\href{https://doi.org/10.1109/TCSS.2023.3241065}{doi:\nolinkurl{10.1109/TCSS.2023.3241065}}


\end{thebibliography}
\end{document}